\documentclass[opre,nonblindrev]{informs_bl}

\usepackage{my_or_definitions}
\usepackage{amsmath}
\usepackage{latexsym}
\usepackage{amssymb}
\usepackage{mathrsfs}
\usepackage{amsfonts}
\usepackage{subfigure}
\usepackage{tabularx}
\usepackage{multirow}
\usepackage{comment}
\usepackage{bigints}
\usepackage[ruled,vlined]{algorithm2e}
\usepackage{algpseudocode}
\usepackage{hyperref}
\usepackage{setspace}
\usepackage{xcolor}
\usepackage{enumitem}
\usepackage{wrapfig}
\hypersetup{
backref=true,
pageanchor=true,
bookmarks,
bookmarksnumbered,
colorlinks=true,
menubordercolor= [rgb]{0, 0, 1},
linkbordercolor= [rgb]{0, 1, 1},
citecolor = [rgb]{0.6, 0, 0},
linkcolor= [rgb]{0, 0, 0.7}
}

\OneAndAHalfSpacedXI 

\usepackage{endnotes}
\let\footnote=\endnote
\let\enotesize=\normalsize
\def\notesname{Endnotes}%
\def\makeenmark{$^{\theenmark}$}
\def\enoteformat{\rightskip0pt\leftskip0pt\parindent=1.75em
  \leavevmode\llap{\theenmark.\enskip}}

\newcommand{\off}{{\textsf{\tiny off}}}
\newcommand{\on}{{\textsf{\tiny on}}}
\newcommand{\onq}{{\textsf{\tiny on-q}}}
\newcommand{\onnq}{{\textsf{\tiny on-nq}}}
\newcommand{\SoS}{{\textsf{SS}}}
\newcommand{\Hyb}{{\textsf{Hyb}}}
\newcommand{\TV}{{\textsf{\tiny TV}}}
\newcommand{\GPG}{{\textsf{GPG}}}
\newcommand{\OPT}{{\textsf{OPT}}}
\newcommand{\SPP}{{\textsf{SPP}}}
\newcommand{\DSPP}{{\textsf{DSPP}}}
\newcommand{\true}{{\textsf{True}}}
\newcommand{\false}{{\textsf{False}}}
\newcommand{\cost}{{\textsf{cost}}}

\newcommand{\Qs}{\mathscr{Q}}
\newcommand{\Xs}{\mathscr{X}}

\newcommand{\Xsb}{\pmb{\mathscr{X}}}
\newcommand{\Qsb}{\pmb{\mathscr{Q}}}

\newcommand{\leqicx}{\leq_{\textit{\tiny icx}}}

\newcommand{\eqst}{=_{\textit{\tiny st}}}

\newcommand{\blue}{}

\graphicspath{{figure/}}

\usepackage[normalem]{ulem}

\usepackage{natbib}
 \bibpunct[, ]{(}{)}{,}{a}{}{,}%
 \def\bibfont{\small}%
 \def\bibsep{\smallskipamount}%
 \def\bibhang{24pt}%

\TheoremsNumberedThrough     
\ECRepeatTheorems

\EquationsNumberedThrough    

\graphicspath{{./figures/}}

\begin{document}


\RUNAUTHOR{Varun Gupta}

\RUNTITLE{Greedy Algorithm for Multiway Matching with Bounded Regret}

\TITLE{Greedy Algorithm for Multiway Matching with Bounded Regret}

\ARTICLEAUTHORS{%
\AUTHOR{Varun Gupta}
\AFF{Booth School of Business, University of Chicago, Chicago, IL 60637, \EMAIL{varun.gupta@chicagobooth.edu}} 
} 

\ABSTRACT{ 
In this paper we prove the efficacy of a simple greedy algorithm for a finite horizon online resource allocation/matching problem, when the corresponding static planning linear program (SPP) exhibits a non-degeneracy condition called the \textit{general position gap} (GPG). The key intuition that we formalize is that the solution of the reward maximizing SPP is the same as a feasibility Linear Program restricted to the optimal basic activities, and under GPG this solution can be tracked with bounded regret by a greedy algorithm, i.e., without the commonly used technique of periodically resolving the SPP. \\
The goal of the decision maker is to combine resources (from a finite set of resource types) into configurations (from a finite set of feasible configurations) where each configuration is specified by the number of resources consumed of each type and a reward. The resources are further subdivided into three types -- \textit{offline} (whose quantity is known and available at time 0), \textit{online-queueable} (which arrive online and can be stored in a buffer), and \textit{online-nonqueueable} (which arrive online and must be matched on arrival or lost).  
Under GPG we prove that,  \textit{(i)} our greedy algorithm gets bounded any-time regret of $\Ocal(1/\epsilon_0)$ for matching reward ($\epsilon_0$ is a measure of the GPG)  when no configuration contains both an online-queueable and an online-nonqueueable resource, and \textit{(ii)} $\Ocal(\log t)$ expected any-time regret otherwise (we also prove a matching lower bound). By considering the three types of resources, our matching framework encompasses several well-studied problems such as dynamic multi-sided matching, network revenue management, online stochastic packing, and multiclass queueing systems.
}%


\KEYWORDS{Dynamic matching, Sum-of-Squares, General Position Gap, Lyapunov analysis, amortized analysis}

\maketitle

%


\section{Introduction}

\paragraph{Problem Setup (informal):} Consider a decision maker who seeks to maximize the total reward by combining resources of different types which arrive online into feasible matching configurations over a horizon of $T$ steps. The set of resource types $\Ical=\{1,\ldots,n\}$ is finite, as is the set of feasible matching configurations $\Mcal=\{1,\ldots,d\}$. The resource consumption matrix $\Mb \in \NN^{n\times d}$ is used to denote the number of units of each resource participating in each match: $M_{im}$ is the number of units of resource type $i$ in configuration $m$. A match of type $m$ generates a reward $r_m$. We assume that for each $i \in \Ical$, the set $\Mcal$ consists of a special configuration consisting exclusively of one unit of resource $i$ with reward 0 -- which can be used to model discarding of resources. 
The resource types are partitioned as $\Ical = \Ical^{\off}\cup \Ical^{\onq} \cup \Ical^{\onnq}$. The set $\Ical^{\off}$ denotes the set of resources available offline to the decision maker; the number of available units of type $i\in\Ical^\off$ are $\lambda_i T$. The set $\Ical^{\on} = \Ical^{\onq} \cup \Ical^{\onnq}$ denotes the set of resources which arrive online \textit{i.i.d.} from a known distribution; let $\lambda_i$ denote the probability that arrival is of type $i \in \Ical^\on$ at any time step (our results will hold a bit more generally, we defer the formal model and results until Section~\ref{sec:model}). Resources in $\Ical^{\onq}$ can be queued on arrival and used to complete a matching configuration at a later time period. Resources in $\Ical^{\onnq}$ can not be queued, and must be matched on arrival or are lost forever (note that a matching configuration can have at most one item from $\Ical^{\onnq}$).\footnote{Our results extend to settings where configurations require multiple units of  resources from $\Ical^\onnq$, but these resources must be committed on arrival to partially completed configurations which have their demand from $\Ical^\off \cup \Ical^\onq$ met. } 

The general framework introduced above with the three types of resources captures many online resource allocation problems studied in the literature:
\begin{enumerate}
    \item {Dynamic Multi-sided Matching} \citep{kerimov2021dynamic,kerimov2021optimality,nazari2019reward}: This model is motivated by kidney matching applications (where resources correspond to donor-recipient pairs and matches correspond to directed cycles where each arc of the cycle corresponds to two pairs where the donor of the tail is compatible with the recipient of the head), ride-pooling application (where the resources correspond to riders and drivers), {\blue and assemble-to-order systems (where resources correspond to orders for assembled products as well as components used in assembly, and matches correspond to an order together with the component parts required to assemble it in some feasible manner)}. In dynamic multi-sided matching all the resources types are in $\Ical^{\onq}$, and the goal is to maximize the matching reward while keeping the queues small in expectation. It is not necessary to match all the arriving resources. Our strongest results are for this case -- we give the first parameter-free algorithm with bounded expected regret for general dynamic multi-sided matching for all times $t \in \{1,\ldots,T\}$.
    \item {Network Revenue Management/Online Knapsack} \citep{jasin2012re, bumpensanti2020re, vera2021online}: A typical example of this application is revenue management in airlines where the offline resources are plane seats on each of the legs operated by the airline. Passengers arrive online with an origin and a destination, and the decision maker offers a route between the origin and destination using legs with available capacity. To map to our setting, we can model the passengers with each origin-destination pair as an online-nonqueueable resource type and the plane seats as offline resources. Feasible configurations consist of one passenger, and a route (set of legs) matching the origin and destination of the passenger.
    \item {Multiclass-multiserver queueing systems with match-dependent rewards} \citep{miller1969queueing,ding2018fluid}: Here the system consists of a pool of $m$ heterogeneous servers, and $n$ customer types. Customers of each type arrive according to a Poisson process. The service time distribution at each server is Exponential with server-dependent mean, however, the utility of a customer-server match depends on the pair. The decision maker makes two sets of decisions: first, which customers to let into the system, and second how to match the customers and servers. To map to our setting, the customers represent online-queueable resources, and the servers (or service opportunities) represent online-nonqueueable resources. Feasible configurations are all customer-server pairs, and are associated with the corresponding matching reward values.
    \item {Online Stochastic bin packing} \citep{csirik2006sum,gupta2020interior, banerjee2020uniform}: Here the decision maker has access to an infinite collection of bins of integer size $B$. Items with scalar (integer) size arrive online and must be packed irrevocably into a feasible bin on arrival. The matching configurations  correspond to multisets of resources with total size at most $B$, and are downward closed (any strict subset of a matching configuration is also a feasible matching configuration). We can view the items as online-queueable resources, however there are a few differences with the framework described above \textit{(i)} this is a cost minimization setting with a cost of 1 per bin, \textit{(ii)} we must pack every arrival in some bin, and \textit{(iii)} the items must be packed on arrival. The first can be addressed by associating a reward of $-1$ with each configuration. The second can be incorporated by requiring that each item must be packed in some configuration (our proposed algorithm already enforces this by associating rejected items with singleton configurations which in this case can be assigned a reward of $-2$, and easily extends to setting where rejections are forbidden as long as downward closed condition for feasible configurations is satisfied). The third requirement is also met by our algorithm because it in fact commits items to configurations at the time of arrival. 
\end{enumerate}

\vspace{3mm}
\paragraph{Static Planning Problem and General Position Gap:} The following Linear Program which finds the reward maximizing matching rates for the deterministic approximation of the online problem is called the Static Planning Linear Program ($\SPP$). {\blue In the $\SPP$, we relax the online aspect of the problem by assuming that all the resources can be indefinitely queued and matched at any time.} The decision variable $x_m$ corresponds to the fractional number of matches of configuration $m$ per time step.
\begin{align}
    \tag{\SPP($\lambdab$)}
    \label{mp:SPP}
    \max_{\{x_m\}_{m\in \Mcal}} : & \sum_{m \in \Mcal} r_m \cdot x_m \\
    \nonumber
    \mbox{subject to} \\
    \nonumber
    \forall i \in \Ical : & \sum_{m \in \Mcal} M_{im} \cdot x_m = \lambda_i, \\
    & x_m \geq 0.
    \nonumber
\end{align}
We denote the optimal value of $\SPP(\lambdab)$ by $\OPT(\lambdab)$. 
Recall that the set $\Mcal$ includes configurations used to model discarding of resources, which is why the resource availability constraints are modeled as equality and not $\leq$. For an optimal basic feasible solution, we will call configurations $m$ with $x_m > 0$ as \textit{basic activities}. We say that $\lambdab$ satisfies the \textit{general position gap} (GPG) condition if 
{\blue there exists some $\epsilon > 0$, such that for an optimal basic feasible solution $\xb^*$ to \SPP($\lambdab$), 
for all $\hat{\lambdab}$ satisfying $|\hat{\lambdab} - \lambdab|_1 \leq \epsilon$ there also exists an optimal basic feasible solution to \SPP($\hat{\lambdab}$) with the same basic activities (non-zero components) as $\xb^*$. We present a more detailed discussion in Section~\ref{sec:model}.} Algorithmically and analytically, GPG allows proving a drift bound on the potential function of interest giving bounded regret.

\vspace{3mm}
\paragraph{Sum-of-Squares algorithm:} Our greedy algorithm is inspired by the Sum-of-Squares algorithm of \cite{csirik2006sum} for the Online Stochastic bin packing problem we described previously. The algorithm of \cite{csirik2006sum} places items greedily so as to minimize the Sum-of-Squares potential function $\Phi_{\SoS}=\sum_{h=1}^{B-1} Q(h)^2$, where $Q(h)$ denotes the number of bins with $h$ units of volume occupied (called level $h$ bins). {\blue The authors prove that their algorithm gets a regret of $\Ocal(\sqrt{T})$ for a special class of distributions called \textit{Perfectly Packable} distributions, while being agnostic to the distribution. However, their algorithm gets $\Ocal(\log T)$ regret compared to the hindsight optimal if the item size distribution is what is called a \textit{Bounded Waste} distribution. Furthermore, the regret improves to $\Ocal(1)$ if the algorithm knows the support of the item size distribution, and is forbidden from creating any \textit{dead-end} level bins -- these are bins into which no set of items in the support of the distribution can be packed to create a level $B$ bin. Our starting point was the observation  that this result can be generalized much more. Let $\Hcal \subseteq \{1,\ldots, B-1\}$ denote the bin levels which are part of the optimal fractional packing (that is, in the optimal fractional packing, there are non-zero bins for each level $h \in \Hcal$) -- we will call these \textit{basic levels} in analogy with basic activities. The sum-of-squares algorithm gets $\Ocal(\log T)$ regret as long as the Sum-of-Squares potential function is modified to $\Phi_{\SoS} = \sum_{h \notin \Hcal} Q(h)^2$, and
the distribution of item sizes satisfies \textit{general position gap} (GPG) (bounded-waste distributions satisfy GPG, but also many distributions which are not Perfectly Packable satisfy GPG). Furthermore, the regret drops to $\Ocal(1)$ if the algorithm is forbidden from creating bins of levels which are dead-ends for all $h \in \Hcal$.} At a high level, the intuition is that we can think of the optimal bin packing as the solution to a network flow feasibility problem where the source node is the empty configuration, the sink nodes correpond to $\Hcal$, and every other node is an internal node where we must have flow balance (which means in the optimal solution there are no bins for levels corresponding to the internal nodes). Minimizing the potential function $\sum_{h \notin \Hcal} Q(h)^2$  ensures that the expected flow-imbalance at internal nodes is bounded in expectation under GPG. 
The algorithm we propose is an adaptation of this observation to the multiway matching setting, where we view each match as a bin configuration. 
Our algorithm commits items on arrival to one of the basic activities of \SPP($\lambdab$) irrevocably and greedily using a  carefully chosen potential function. In addition, we treat offline resources like online resources by injecting them at a regular schedule for bookkeeping purposes; for creating matches we allow using offline resources which have not ``arrived'' yet. A side effect is that the arrival distribution for our algorithm is non-stationary. We therefore prove our regret results under a more general non-stationary variant of the online matching problem combining Lyapunov drift analysis with techniques from amortized analysis of online algorithms.

\vspace{3mm}

\paragraph{Prior Work:} We briefly mention the strength and weaknesses of our results in relation to the existing literature, but do not make an attempt to provide an exhaustive review of the results (for which we refer the reader to the recent papers \citep{vera2021online} and \citep{kerimov2021dynamic}). The main takeaway message of our paper is that the knowledge of basic feasible activities of the optimal offline solution is immensely powerful, and can be used to create quite simple greedy algorithms where the precise values of {\blue arrival rates $\lambdab$  and rewards $\rb$} are immaterial and small fluctuations and non-stationarity around $\lambdab$ impact the regret gracefully. While there have been some papers which implicitly exploit this intuition, we have not seen it formalized, and not in the generality considered in the present paper. We have already mentioned that \cite{csirik2006sum} prove bounded regret of Sum-of-Squares algorithm for online stochastic bin packing for a special case where effectively they are using a greedy algorithm restricted to optimal basic activities, but have not observed the general scheme to the best of our knowledge. Recently, \cite{kerimov2021optimality} propose a `greedy' algorithm with bounded regret for \textit{bipartite} version of multiway matching under GPG. The notion of `greedy' employed in \cite{kerimov2021optimality} is that if at the time of the arrival the item can be matched to another queued item using an optimal basic feasible configuration, then this match is performed. Our results show the right generalization of `greedy' beyond bipartite matching -- the items should be committed to a matching configuration on arrival even if a match can not be completed at that time. 

While our results are the state-of-the-art for multiway matching under GPG, they are not state-of-the-art for Network Revenue Management and Online Stochastic Bin Packing where there are several results which get bounded regret for these problems \textit{without GPG}, for example, \citep{arlotto2019uniformly,bumpensanti2020re, vera2021online, banerjee2020uniform}. These results are based on frequently resolving the SPP, which we prove to be unnecessary under GPG. {\blue However we would like to point out a few differences in the flavor of results. \textit{(i)} The benchmark against which these papers get $\Ocal(1)$ regret is the expected hindsight optimal of the sample path (i.e., $\expct{\OPT}$), whereas the regret guarantee in this paper also holds against the fluid optimal $\OPT(\lambdab) = \OPT(\mathbf{E})$. When GPG condition does not hold, in general $\OPT(\lambdab)-\expct{\OPT} = \Omega(\sqrt{T})$. Therefore, the fluid benchmark is too strong and even hindsight optimal can not get $\Ocal(1)$ regret against it. When GPG does hold, the expected hindsight optimal and fluid optimal are $\Ocal(1)$ apart. \textit{(ii)} All these results only get bounded regret at the terminal time $T$. In fact as \cite{banerjee2020uniform} show, knowledge of the horizon $T$ is critical for the $\Ocal(1)$ regret guarantee to hold. In contrast, for stochastic bin packing, we prove bounded anytime regret without the knowledge of $T$ under GPG. If GPG holds, we believe the algorithms of \citep{banerjee2020uniform,bumpensanti2020re} still would not get $\Ocal(1)$ anytime regret since they only resolve the SPP sporadically, whereas \citep{vera2021online} would get $\Ocal(1)$ regret since it resolves a planning problem at every time step.}

{\blue The General Position Gap assumption we make is equivalent to a dual non-degeneracy condition which requires that the solution to a partial dual of \eqref{mp:SPP} be unique. This dual non-degeneracy appears as a condition for small regret in many online resource allocation problems (e.g., \citep{huang2009delay,jasin2012re,chen2021fairer}). For example, \cite{huang2009delay} study an online bandwidth allocation problem, where the evolution of the queues corresponds to a stochastic ascent algorithm for the dual function. Here dual non-degeneracy leads to the dual function being locally polyhedral around its unique optimal, which in turn gives concentration of the duals. The analysis in the current paper uses a tailor made Lyapunov function, and not the dual function. However, the spirit of the drift arguments for the Lyapunov function and the dual function are similar.}

The knowledge of the basic feasible activities has been exploited in control of multiclass-multiserver queueing systems with match-dependent service rates \citep{stolyar2013tightness}. However, in their work the emphasis is on minimizing delay under a many-servers heavy-traffic regime (there are no match-dependent rewards). The authors devise a static priority policy based on the tree induced by the basic feasible activities which is oblivious to the arrival rates ($\lambda_i$ in our setup for customers) as well as service rates (the matrix $\Mb$ in our setup). \cite{kerimov2021dynamic} also propose a priority policy for the case of bipartite matching based on the basic feasible activities: whether priority policies beyond the bipartite case are possible is an open research question.

\vspace{3mm}
\paragraph{Summary of Contributions:}
\begin{itemize}
    \item We introduce a model of multiway matching that unifies the existing work on dynamic multi-sided matching, network revenue management, and queueing network control, among others, via the introduction of three types of resources -- offline, online-queueable, and online-nonqueueable.
    \item We introduce a new greedy algorithm for the multiway matching problem, and formalize the intuition behind the optimality of greedy algorithms which only restrict their actions to the optimal basic feasible activities of the static planning problem. Our greedy algorithm has $\Ocal(1/\epsilon_0)$ any-time regret when no configuration has both an online-queueable and online-nonqueueable resources (e.g., in dynamic multisided matching, network revenue management) and tight $\Ocal(\log t)$ regret otherwise (e.g., queueing network control).
    \item We present a new regret analysis which combines Lyapunov function with amortized analysis to handle non-stationary arrival sequences, and should be of independent interest.
\end{itemize}

\vspace{3mm}
\paragraph{Notation:}
We use the notation $[i]=\{1,\ldots,i\}$. The notation $[s,t]$ is used for the discrete interval $\{s,s+1, \ldots, t\}$ as well as for the continuous interval $[s,t]$, where the use should be clear from the context. We use bold letters to denote vectors and matrices. For an $n\times d$ matrix $\Mb$, we use $\Mb_j$ to denote its $j$th column, and $\Mb_{\Jcal}$ for $\Jcal \subseteq [d]$ to denote the submatrix of $\Mb$ obtained by selecting the columns in $\Jcal$. Similarly, for a vector $\xb$, $\xb_{\Jcal}$ denotes the vector obtained by selecting the components of $\xb$ corresponding to the subset $\Jcal$.
$\eb_i$ denotes a column vector with $i$th entry 1 and the rest 0. 
For quantities which are functions of time and for stochastic processes, we use superscript $t$ as the time index. For any process $o$, we use $\bar{o}^{[t_1,t_2]} := \frac{1}{t_2-t_1+1}\sum_{t \in [t_1,t_2]}o^t$. The expression $(x)^+ = \max\{x,0\}$ and $(x)^- = \max\{ -x, 0\}$. For two probability distributions $\pb,\qb$, we denote the total variation distance by $
 \norm{\pb-\qb}_{\TV} = \frac{1}{2}|\pb - \qb|_1$, and the total variation ball around $\pb$ of radius $\delta$ is defined as $\Bcal_{\epsilon,\TV}(\pb) = \{ \qb \ |\  \norm{\pb-\qb}_\TV \leq \epsilon \}$. For an event $\Ecal$, $\ind\{\Ecal\}$ denotes the indicator random variable. For two vectors $\xb,\yb$, we use $\xb \leq \yb$ to mean $\xb$ is component-wise no larger than $\yb$.

\vspace{3mm}
\paragraph{Outline:} We begin in Section~\ref{sec:iid} by covering the somewhat simpler setting of multiway matching with only online-queueable resources and \textit{i.i.d.} arrivals to convey the main ideas with minimum notational burden. The analysis in this setting largely uses tools already developed in \citep{csirik2006sum}, but the algorithm is different than \citep{csirik2006sum}. In the rest of the paper we focus on general multiway matching instances with all three types of resources (offline, online-queueable, online-nonqueueable), and non-stationary arrivals. To convert a problem with both offline and online resources into one with only online resources, our greedy algorithm is wrapped within a procedure that creates an online arrival sequence for the offline resources. In Section~\ref{sec:emulation} we describe this procedure. In Section~\ref{sec:model} we describe the general online matching model, state the General Position Gap assumptions needed for our results, and state our main results (Theorem~\ref{thm:mainUB} and Theorem~\ref{thm:mainLB}). In Section~\ref{sec:SoS} we present our Sum-of-Squares algorithm, and perform regret analysis in Section~\ref{sec:analysis}. Section~\ref{sec:experiments} ends with some simulation experiments.

{\blue
\section{Prelude - Multiway matching with online-queueable resources and \textit{i.i.d.} arrivals}
\label{sec:iid}

To convey the main ideas that go into the algorithm and regret analysis, in this section we consider the case of multiway matching with only online-queueable resources ($\Ical = \Ical^{\onq} = [n]$) over a horizon of $T$ time steps, with the type of item arriving at time $t \in [T]$ sampled \textit{i.i.d.} from a known distribution $\lambdab$. 
Starting in Section~\ref{sec:emulation}, we study general online matching instances with non-stationary arrivals, and with online-nonqueueable and offline resources, which require some modifying the algorithm and slightly advanced technical analysis.

We assume the set of feasible matches is denoted by the matching matrix $\Mb \in \{0,1\}^{n \times d}$. We will denote the reward of configuration $m$ by $r_m$, the set of items participating in configuration $m$ by $\Ical(m)$, and $\ell_m = |\Ical(m)|$. 
Let $\Mcal_+$ be an optimal basis for the static planning problem \eqref{mp:SPP}, and let $\xb^*$ be the optimal solution corresponding to $\Mcal_+$ ($x^*_m = 0$ for $m \notin \Mcal_+$). The general position gap assumption amounts to saying that for all $\hat{\lambdab}$ in a total variational ball of radius $\epsilon_0$ around $\lambdab$ (i.e.,  $\hat{\lambdab} \in \Bcal_{\epsilon_0,\TV}(\lambdab)$) there is an optimal solution to $\SPP(\hat{\lambdab})$ with basis $\Mcal_+$. 

\paragraph{\bf Algorithm:} As we mentioned in the introduction, the crux of our algorithm is to commit an item to a configuration in $\Mcal_+$ at the time of its arrival to the system, where the identity of this configuration is decided based on greedily minimizing a Sum-of-Squares potential function. Our algorithm maintains a queue for each configuration $m\in \Mcal_+$, and each resource $i\in \Ical(m)$, where this queue denotes the set of items of resource $i$ that have been committed to configuration $m$ but not converted into a match yet. Let $Q^{t}_{i,m}$ denote the queue length for the pair $(i,m)$ just before the arrival at time $t+1$. Let $\Qb^t_m = (Q^t_{i,m})_{i\in \Ical}$ denote the state of the queues for configuration $m$, and $\Qb^t = (\Qb^t_m)_{m \in \Mcal_+}$ denote the state of all the queues. If an item of type $i$ arrives at time $t$ and is committed to configuration $m$, then 
\begin{align*}
    \Qb^t_m &= \begin{cases}
        \Qb^{t-1}_m + \eb_i - \Mb_m & \mbox{if $\Qb^{t-1}_m+\eb_i \geq \Mb_m$,} \\
        \Qb^{t-1}_m + \eb_i & \mbox{otherwise.}
    \end{cases}
\end{align*}
That is, if there are sufficient number of resources in the queues $\Qb_m^{t-1}$ along with the new arrival to complete a match then all the queues decrease accordingly. Otherwise the new arrival is added to the queue $Q_{im}$.

For a configuration $m \in \Mcal_+$, pick any permutation $\sigma_m$ of $\Ical(m)$. Then the cyclic Sum-of-Squares potential function for configuration $m$ as a function of the state $\Qb_m$ is defined as:
\begin{align}
    \Phi_m(\Qb_m) &= \left( Q_{\sigma_m(1),m} - Q_{\sigma_m(\ell_m),m}  \right)^2 + \sum_{k=2}^{\ell_m} \left( Q_{\sigma_m(k),m} - Q_{\sigma_m(k-1),m}  \right)^2,
\end{align}
and the overall Sum-of-Square potential is
\begin{align}
    \Phi(\Qb) &= \sum_{m \in \Mcal_+} \Phi_m(\Qb_m).
\end{align}
Our greedy Sum-of-Squares algorithm commits the arrival at time $t$ to a configuration so as to minimize $\Phi(\Qb^t)$. Intuitively, for each $m$, one of the $Q^t_{im}$ must always be 0 otherwise a match can be completed. By trying to minimize the sum of squared differences, the Sum-of-Squares potential function tries to keep all $Q^t_{im}$ small.

\paragraph{\bf Analysis:} Let $\Ab^t = (A^t_i)_{i \in \Ical}$ denote the number of arrivals of each resource type until and including time $t$. Let $\Xb^t = (X^t_m)_{m\in \Mcal_+}$ denote the number of completed matched of configurations in $\Mcal_+$ until and including time $t$. The next lemma bounds the expected regret of any algorithm which commits items to configurations in $\Mcal_+$ on arrival in terms of the queue lengths $\Qb^t$. This lemma does not require general position gap assumption, however subsequent analysis then shows that $\expct{Q^t_{im}}$ is bounded under the greedy Sum-of-Squares based algorithm if the general position gap assumption holds.  

\begin{lemma}
\label{lem:regret_decomposition}
For multiway matching with online-queueable resources and \textit{i.i.d.} arrivals, the expected regret at time $t$ for any algorithms that commits items to configurations in $\Mcal_+$ on arrival is bounded as follows:
\begin{align*}
    \expct{ \sum_{m \in \Mcal_+} r_m \cdot X^t_m } & \geq \expct{\OPT(\Ab^t)}-  \rb_{\Mcal_+} \Mb_{\Mcal_+}^{-1} \expct{\sum_{m \in \Mcal_+} \Qb^t_m}.
\end{align*}
\end{lemma}
\begin{proof}{Proof:}
Since $\Mcal_+$ is a basis, $\Mb_{\Mcal_+}$ is full rank. And since we assume that the algorithm only commits items to configurations in $\Mcal_+$, $\Xb^t$ is given by:
\[  \Xb^t_{\Mcal_+} = \Mb_{\Mcal_+}^{-1} \left( \Ab^t - \sum_{m\in \Mcal_+} \Qb^t_m \right).\]
Further, $\expct{\Ab^t}=\lambdab \cdot t$ implies $\OPT(\expct{\Ab^t})=\rb_{\Mcal_+}^\top \Mb_{\Mcal_+}^{-1} \expct{\Ab^t}$.
Therefore, 
\begin{align*}
    \expct{\sum_{m\in \Mcal_+} r_m \cdot X^t_m} &= \expct{\rb_{\Mcal_+}^\top \Mb_{\Mcal_+}^{-1} \left( \Ab^t - \sum_{m \in \Mcal_+} \Qb^t_m \right)} \\
    &= \rb_{\Mcal_+}^\top \Mb_{\Mcal_+}^{-1} \expct{\Ab^t}  - \rb_{\Mcal_+}^\top \Mb_{\Mcal_+}^{-1} \expct{\sum_{m \in \Mcal_+} \Qb^t_m} \\
    & = \OPT(\expct{\Ab^t})  - \rb_{\Mcal_+}^\top \Mb_{\Mcal_+}^{-1} \expct{\sum_{m \in \Mcal_+} \Qb^t_m }\\
    & \geq \expct{\OPT(\Ab^t)}  - \rb_{\Mcal_+}^\top \Mb_{\Mcal_+}^{-1} \expct{\sum_{m \in \Mcal_+} \Qb^t_m }.
\end{align*}
The last step follows since $\OPT(\cdot)$ is concave and piece-wise linear (this fact is apparent by looking at the dual of \eqref{mp:SPP} which is given in \eqref{mp:DSPP}). 
\hfill\BlackBox
\end{proof}

 The next lemma states that if $\Phi(\Qb^{t})$ is large and the General Position Gap condition holds, then in expectation this function has a large negative drift at time $t+1$. The proof of this lemma is almost identical to Theorem 3.4 from \citep{csirik2006sum} where authors prove a similar \textit{Expected Decrease Condition} for stochastic bin packing with bounded waste distributions. The expected decrease condition is also central for our analysis. 
\begin{lemma} 
\label{lem:drift_main_illustration}
The Lyapunov function $\Psi^t := \sqrt{\Phi(\Qb^t)}$ satisfies the following conditions:
\begin{enumerate}[label=(\alph*)]
\item Bounded variation:  $\left| \Psi^{t+1} - \Psi^t \right| \leq \sqrt{2}$ with probability 1
\item Expected Decrease: Under the General Position Gap condition with parameter $\epsilon_0$ 
\[ \expct{\Phi(\Qb^{t+1}) - \Phi(\Qb^t) \left| \Phi(\Qb^t) \geq \frac{16n^4}{\epsilon^2_0} \right.} \leq - \frac{\epsilon_0 \sqrt{\Phi(\Qb^t)}}{2 n^2}, \]
which further gives,
\[ \expct{\Psi^{t+1} - \Psi^t | \Psi^t \geq \frac{4n^2}{\epsilon_0}} \leq - \frac{\epsilon_0 }{4n^2}. \]
\end{enumerate}
\end{lemma}
\begin{proof}{Proof:}
\textit{(a)} Let $m\in \Mcal_+$ be the configuration to which a resource is committed at time $t$. Without loss of generality assume $\Ical(m)=\{1,\ldots, \ell\}$, let $i$ be the resource added, and let the permutation $\sigma_m$ used to define $\Phi_m(\Qb_m)$ be the identity permutation, so that
\[ \Phi_m(\Qb_m) = \left(Q_{1,m}-Q_{\ell,m}\right)^2 + \sum_{k=2}^\ell \left( Q_{k,m} -Q_{k-1,m} \right)^2.\]
Assume $\Phi_m(\Qb^{t+1}_m) \geq \Phi_m(\Qb^{t}_m)$; the other case is handled in a similar manner. In this case examining the queue length differences involved in $\Phi_m$,
\[ Q^{t+1}_{i,m}-Q^{t+1}_{i-1,m} = Q^{t}_{i,m}-Q^{t}_{i-1,m}+1 , \qquad Q^{t+1}_{i+1,m}-Q^{t+1}_{i,m} = Q^{t}_{i+1,m}-Q^{t}_{i,m}-1,  \]
and all the other differences are unchanged. Therefore, for some 
for $x,y \geq 0$,,
\[ \Phi_m(\Qb^{t+1}_m) \leq \Phi_m(\Qb^t_m) + (x+1)^2 +(y+1)^2 - x^2-y^2.\]
Now, following the proof of Lemma 3.5 of \cite{csirik2006sum}:
\begin{align*}
 \sqrt{\Phi(\Qb^{t+1})} - \sqrt{\Phi(\Qb^t) }  &\leq  \sqrt{ \Phi(\Qb^t) + (x+1)^2+(y+1)^2 - x^2 - y^2 } - \sqrt{\Phi(\Qb^t) } \\
 &= \frac{2x+2y+2 }{\sqrt{\Phi(\Qb^t) +(x+1)^2+(y+1)^2 - x^2 - y^2 } + \sqrt{\Phi(\Qb^t)} } \\
 & \leq \frac{2x+2y+2 }{\sqrt{(x+1)^2+(y+1)^2} + \sqrt{x^2 + y^2} } \\
 \intertext{which is maximized at $x=y=(x+y)/2$ for a fixed value of $x+y=z$, giving the upper bound,}
 & \leq \frac{2z+2 }{\sqrt{\frac{z^2}{2}+2z+2}+\sqrt{\frac{z^2}{2}}} \\
 & = \sqrt{2}.
\end{align*}
\textit{(b)} The proof involves two steps. The first step involves showing that if $\Phi(\Qb^t) \geq 4n^4$, then there exists some resource type $i^* \in \Ical$ such that if an item of type $i^*$ is committed by our greedy algorithm, then the resulting state satisfies 
\[ \Phi(\Qb^{t+1})-\Phi(\Qb^t)  \leq - \frac{\sqrt{\Phi(\Qb^t)}}{n^2} .\]
This statement is proved as Lemma~\ref{lem:negative_drift} in the Appendix, and its proof follows similar to Lemma 3.7 of \citep{csirik2006sum}. 

The second step involves showing that for all distributions $\hat{\lambdab} \in \Bcal_{\epsilon_0,\TV}({\lambdab})$, there exists a randomized commitment policy $\Rcal(\hat{\lambdab})$ (formally defined later in Definition~\ref{def:Rlambda}), such that if an item sampled from $\hat{\lambdab}$ is committed to one of the configurations in $\Mcal_+$ by $\Rcal(\hat{\lambdab})$ at time $t$ then the expected change in the Sum-of-Squares potential is:
\[ \expct{\Phi(\Qb^{t+1})}-\Phi(\Qb^t) = 2 .\]
This statement is proved as Lemma~\ref{lem:expected_inc} in the Appendix, and its proof follows similar to Lemma 2.2 of \citep{csirik2006sum}. Since our algorithm minimizes $\Phi(\Qb^{t+1})$ given $\Phi(\Qb^t)$, this gives an upper bound on $\expct{\Phi(\Qb^{t+1})}-\Phi(\Qb^t)$ for our Sum-of-Squares based greedy algorithm.

Now, suppose the state at time $t$ satisfies $\Phi(\Qb^t) \geq \frac{16n^4}{\epsilon_0^2} \geq 4n^4$, and let $i^*$ be the item from the first step above which leads to a large negative drift. We can decompose $\lambdab$ as:
\[ \lambdab  = \epsilon_0 \cdot \eb_{i^*} + (1-\epsilon_0) \cdot  \hat{\lambdab} \]
where $\hat{\lambdab} = \frac{\lambdab - \epsilon_0 \cdot \eb_i}{1-\epsilon_0} \in \Bcal_{\epsilon_0,\TV}(\lambdab)$. Combining the above two steps, the expected drift in $\Phi$ for the greedy algorithm is upper bounded by:
\begin{align*}
    \expct{\Phi(\Qb^{t+1})} - \Phi(\Qb^t) &\leq - \epsilon_0  \cdot \frac{\sqrt{\Phi(\Qb^t)}}{n^2} + (1-\epsilon_0) \cdot 2 \\
    &\leq - \epsilon_0  \cdot \frac{\sqrt{\Phi(\Qb^t)}}{n^2} + \epsilon_0  \cdot \frac{\sqrt{\Phi(\Qb^t)}}{2n^2} \\
    \intertext{(since $\Phi(\Qb^t) \geq \frac{16n^4}{\epsilon_0^2}$ implies $2 \leq \epsilon_0  \cdot \frac{\sqrt{\Phi(\Qb^t)}}{2n^2}$)}
    &= - \epsilon_0  \cdot \frac{\sqrt{\Phi(\Qb^t)}}{2n^2}.
\end{align*}

Using the inequality (Lemma 3.6 of \citep{csirik2006sum}): for $x,y \geq 0$,
\[ x - y \leq \frac{x^2-y^2}{2y},\]
we get,
\begin{align*}
    \expct{\Psi^{t+1} - \Psi^t \left| \Qb^t;\Psi^t \geq \frac{4n^2}{\epsilon_0} \right.} \leq \frac{\expct{\Phi(\Qb^{t+1}) - \Phi(\Qb^t) \left.| \Qb^t; \Psi^t \geq \frac{4n^2}{\epsilon_0}\right.}}{2 \sqrt{\Phi(\Qb^t)}} \leq - \epsilon_0  \cdot \frac{\sqrt{\Phi(\Qb^t)}}{4n^2 \sqrt{\Phi(\Qb^t)}} = -\frac{\epsilon_0}{4n^2}.
\end{align*}
These final calculations are the essence of Lemma~\ref{lem:negative_drift_main} which requires a little bit of notational overhead. We reproduced them here to keep this section self-contained to the extent possible.   
\hfill\BlackBox
\end{proof}

In \citep{csirik2006sum}, after proving the expected decrease condition, the authors apply a result of \cite{hajek1982hitting} to conclude that $\expct{\Psi^t}$ and hence regret is $\Ocal(1)$. However applying the result of \cite{hajek1982hitting} gives a regret bound that depends on the General Position Gap parameter as $\Ocal(1/\epsilon_0^2)$ and not $\Ocal(1/\epsilon_0)$. To get a better regret we ``upper bound'' the process $\Psi^t$ by a tractable reflected random walk with \textit{i.i.d.} increments. We begin with the definition of multivariate increasing convex order from \citep[Definition 3.4.1, Theorem 3.4.2]{muller2002comparison}, which is the sense in which we will upper bound the process $\{\Psi^t\}_{t\in [T]}$.

\begin{definition}[Increasing Convex Order]
\label{defn:icx}
Let $\Xb$ and $\Yb$ be $n$-dimensional random vectors with finite expectations. Then $\Xb$ is less than $\Yb$ in increasing convex order ($\Xb \leqicx \Yb$) if $\expct{f(\Xb)}\leq \expct{f(\Yb)}$ for all increasing convex functions $f:\RR^n \to \RR$ such that the expectations exist. Also, $\Xb \leqicx \Yb$ if and only if there are random vectors $\hat{\Xb} \eqst \Xb$ and $\hat{\Yb} \eqst \Yb$ such that $\expct{\hat{\Yb} | \hat{\Xb}} \geq \hat{\Xb}$ almost surely.
\end{definition}

{\blue
\begin{lemma}
\label{lem:icx_bound}
Let $\{\Upsilon^t\}$ be an $\{\Fcal^t\}$-adapted stochastic process satisfying
\begin{enumerate}[label=(\alph*)]
\item Bounded variation: $|\Upsilon^{t+1} - \Upsilon^t| \leq K_1$ with probability 1,
\item Expected Decrease: $\expct{\Upsilon^{t+1} - \Upsilon^t | \Fcal_t; \Upsilon^t \geq K_2} \leq - 2 \delta' K_1$. 
\end{enumerate}
Let $\{\Gamma^{t}\}$ be a reflected random walk with step size $K_1$, reflection boundary $\Gamma_{\min} := K_1 \left(1+\ceil{K_2/K_1}\right) $, initial condition $\Gamma^{\tau} = \max\{ \Gamma_{\min} , K_1 \ceil{\Upsilon^\tau/K_1}\} $, and
\[ \Gamma^{t+1} = \max\{  \Gamma^t + \xi^{t+1} K_1 , \Gamma_{\min} \},  \ t = \tau,\ldots, T-1,\]
where $\xi^{t} \in \{-1,+1\}$ are \textit{i.i.d.} random variables taking the value $+1$ with probability $\frac{1}{2} - \delta'$ and $-1$ otherwise. Then,
\[  \left\{ \Upsilon^t\right\}_{t\in [\tau,T]}  \leqicx \left\{ \Gamma^t\right\}_{t\in [\tau,T]}.  \]
\end{lemma}
}

We now combine Lemma~\ref{lem:drift_main_illustration} and Lemma~\ref{lem:icx_bound} to prove the main regret result for this Section.

\begin{theorem}
\label{thm:regret_illustration}
For all $t \geq 1$, the Lyapunov function $\Psi^t = \sqrt{\Phi(\Qb^t)}$ is upper bounded as 
\[ \expct{\Psi^t} \leq  \frac{8n^2}{\epsilon_0} + \sqrt{2}. \]
Consequently, the regret for the multiway matching problem at time $t$ is bounded as:
\begin{align*}
    \expct{ \sum_{m \in \Mcal_+} r_m \cdot X^t_m } & \geq \expct{\OPT(\Ab^t)}-  \Ocal\left(\frac{n^3}{\epsilon_0}\right).
\end{align*}
\end{theorem}
\begin{proof}{Proof:} 
By  Lemma~\ref{lem:drift_main_illustration} and Lemma~\ref{lem:icx_bound}, we can upper bound the stochastic process $\Psi^t$ in increasing convex order by a random walk with \textit{i.i.d.} increments in $\pm \sqrt{2}$, expected increment $-\frac{\epsilon_0}{4n^2}$, and reflection boundary $\frac{4n^2}{\epsilon_0}+\sqrt{2}$. This gives the upper bound on $\expct{\Psi^t}$ mentioned in the theorem statement since expectation is an increasing convex function. Next, we have the bound,
\[  \sum_{m\in \Mcal_+} \max_{i \in \Ical(m)} Q^t_{im} \leq n \Psi^t,\]
and from Lemma~\ref{lem:regret_decomposition},
\begin{align*}
    \expct{ \sum_{m \in \Mcal_+} r_m \cdot X^t_m } & \geq \expct{\OPT(\Ab^t)}-  \rb_{\Mcal_+} \Mb_{\Mcal_+}^{-1} \expct{\sum_{m \in \Mcal_+} \Qb^t_m} \\
    & \geq \expct{\OPT(\Ab^t)} - \sum_{m \in \Mcal_+} r_m \max_{i \in \Ical(m)} Q^t_{im} \\
    & \geq \expct{\OPT(\Ab^t)} - \max_{m\in \Mcal_+} r_m \cdot \sum_{m \in \Mcal_+}  \max_{i \in \Ical(m)} Q^t_{im} \\
    & \geq \expct{\OPT(\Ab^t)} - \Ocal\left( \frac{n^3}{\epsilon_0} \right). 
\end{align*}
\hfill\BlackBox
\end{proof}

{\bf Peeking ahead:} When we consider multiway matching with offline and online-nonqueueable resources in the sequel, one main difference that occurs is that we need to introduce a notion of \textit{virtual queues} in addition to the true queues $Q^t_{im}$ we used in this section. It is the virtual queues that are used by the Sum-of-Squares based greedy algorithm to make its decisions. A second difference is that to handle offline resources in a similar manner to online resources, we inject them at a constant rate into the system. This leads to a non-stationary arrival process, which requires studying a different Lyapunov function than $\Psi^t$, and blending drift analysis with elements of amortized analysis since the expected decrease condition does not hold for $\Psi^t$ anymore. 
}

\section{Intermezzo - Emulating Online Arrivals for Offline Resources}
\label{sec:emulation}

To treat offline resources in a similar manner to online resources in our greedy algorithm, we will emulate an online arrival process for them {\blue by injecting resources at a ``constant rate'' while at the same time ensuring that the total number of offline resources injected of each type equals their initial inventory}. In this section we  describe this process, which creates an instance with a non-stationary arrival process. As a result, the rest of this paper will study online matching under a somewhat general non-stationary arrival process.

Let $\Ical^\off = [n^\off]$, and $\Ical^\on = [n^\off+1,n]$. 
Let $T'$ denote the horizon for the original matching problem, and let $\lambda_i T' \in \NN$ for $i \in \Ical^\off$ denote the number of units of offline resource $i$ available initially. Let $\{\lambda_i\}_{i \in \Ical^\on}$ denote the probability distribution for the online resources. That is, for each $t \in [T']$, the type of the arriving item is sampled \textit{i.i.d.} from $\{\lambda_i\}_{i \in \Ical^\on}$. Let $\lambdab^\on \in \RR^n$ be such that $\lambda^\on_i=0$ for $i \in \Ical^\off$ and $\lambda^\on_i=\lambda_i$ otherwise.

From the above primitives, we create another online matching instance of \textit{effective horizon} $T = T'\left( 1 + \sum_{i \in \Ical^\off} \lambda_i \right)$ where at any time the arrival is either deterministic and of a type in $\Ical^\off$, or sampled according to $\lambdab^\on$. Define $\bar{\lambda}_i = \lambda_i \cdot T'/T$ for all $i \in \Ical$ as the mean rate of resource of type $i$, and $\bar{\lambda}_0 = \sum_{i \in \Ical^\on} \bar{\lambda}_i$ as the aggregate mean arrival rate of online resources. The procedure is described in Algorithm~\ref{alg:emulation}.

\begin{algorithm}[h!]
\small
\DontPrintSemicolon
   \caption{Virtual online arrival process for offline resources}
   \label{alg:emulation}
 \SetKw{KwDef}{Definition:} 
 \SetKw{KwInit}{Initialize:} 
 \SetAlgoLined
 \KwInit{} $A^0_i = 0$ for each $i \in \Ical^\off \cup \{0\}$
 
 \For{$t=1, \ldots, T$}{
     Let $i = \argmin\left\{i \in \Ical^\off \cup \{0\} \ : \ A^{t-1}_i - \bar{\lambda}_i\cdot (t-1) \right\}$  \tcp*[r]{Break ties arbitrarily}
 \eIf{$ i=0$}{
     $\lambdab^t = \lambdab^\on$
 }
 {
     $\lambdab^t = \eb_i$
 }
     $A^{t}_i = A^{t-1}_i + 1$
 }
\end{algorithm}

{\blue In words, the emulation algorithm looks at the deficit of the total number of resources of type $i$ injected by time $t-1$, $A_i^{t-1}$ ($i=0$ for online, otherwise there is a separate type for each resource in $\Ical^\off$), compared to the expectation, $\bar{\lambda}_i \cdot (t-1)$, and injects the type that is most in deficit. } The following lemma states that the discrepancy between the non-stationary instance $\{\lambdab^t\}_{t \in [T]}$ created by Algorithm~\ref{alg:emulation} and the stationary instance $\bar{\lambdab}$ is bounded at all times.
\begin{lemma}
\label{lem:bounded_discrepancy}
For the emulation in Algorithm~\ref{alg:emulation}, for each $i\in \Ical^\off \cup \{0\}$;  $ -n^\off \leq A_i^t - \bar{\lambda}_i \cdot t   \leq  1$ for each $t \in [T]$, and $A_i^T  = \bar{\lambda}_i \cdot T$. As a consequence, letting $\tau = (n^\off+1)/\lambda_{\min}$, for all $t \in [\tau+1,T]$,
\[ \bar{\lambdab} \cdot \left( t - 2 \tau \right) \leq  \sum_{s=\tau+1}^t \lambdab^s \leq \bar{\lambdab} \cdot t \]
where $\bar{\lambda}_{\min} := \min_{i \in \Ical} \bar{\lambda}_i$; and $ \sum_{s=1}^T \lambdab^s = \bar{\lambdab} \cdot T$.
\end{lemma}

{\blue In the next section we will define a multiway matching model with a somewhat general non-stationary arrival process satisfying a smoothed version of the general position gap assumption (Assumption~\ref{assum:smoothed}) which we call Smoothed $\GPG_{\delta, \tau}$, and prove regret result for such non-stationary instances in Theorem~\ref{thm:mainUB}. Lemma~\ref{lem:bounded_discrepancy} goes together with Assumption~\ref{assum:smoothed} by showing that if the arrival process of online items is stationary and satisfies GPG with parameter $\epsilon_0$, then the non-stationary instance created by Algorithm~\ref{alg:emulation} satisfies $
\GPG_{\delta,\tau}$ with $\tau = (n^\off+1)/\lambda_{\min}$ and $\delta = \epsilon_0$.}

\section{Main act - Model and Results}
\label{sec:model}

{\bf Resource types and matching configurations:} We consider an online matching instance with a horizon of $T$ time steps.
The resource types are denoted by $\Ical = [n]$ which are further partitioned into offline resources: $\Ical^\off = [n^\off]$, online-queueable: $\Ical^\onq = [n^\off+1, n^\off+n^\onq]$, and online-nonqueueable: $\Ical^\onnq = [n^\off+n^\onq+1,n]$. The set of feasible matches is denoted by the matching matrix $\Mb \in \NN^{n\times d}$ with $M_{im}=1$ if resource $i$ participates in $m$. Our algorithm and analysis extends to general $M_{im}\in \NN$ or $M_{im} \in \RR_+$ to capture different amounts of resource $i$ needed by appropriately modifying the potential function, but to keep exposition clear in this paper we focus on $M_{im}\in \{0,1\}$. We will use the notation $\Ical(m)$ to denote the set of resources that participate in configuration $m$, and $i \in m$ as shorthand for $i \in \Ical(m)$. Let $\ell_m = |\Ical(m)|=\sum_i M_{im}$ be the number of distinct resource types participating in $m$. A match of type $m$ generates a reward $r_m$. To model the possibility of discarding resources, we assume that for each $i \in \Ical$, there is a configuration with resource consumption vector $\eb_i$ and reward 0.\footnote{Our results extend to the case where discarding is prohibited as long as the set of configurations is downward closed, that is, strict subsets of feasible configurations are also feasible.}

\vspace{3mm}

{\bf Arrival process:} {\blue We will denote the type of the arriving
resource at time $t$ by $I^t \in \Ical$, and the distribution from which the type $I^t$ is sampled as $\lambdab^t$. We will use $\{\Fcal^t\}_{t\in [T]}$ to denote the filtration generated by the stochastic process $\{I^t\}_{t \in [T]}$, and assume that $\{\lambdab^t\}_{t \in [T]}$ is an $\{\Fcal^{t}\}_{t\in [T]}$-predictable process. That is, the distribution  $\lambdab^t$ from which the type $I^t$ is sampled is itself a random quantity, which is $\Fcal^{t-1}$-measurable. Conditioned on the sequence $\{\lambdab^t\}_{t\in [T]}$, $\{I^t\}_{t\in [T]}$ are assumed to be mutually independent.
Define $Z^t_i = \ind\{I^t=i\}$, and  
$A^t_i = \sum_{s=1}^t Z^s_i$ as the cumulative number of arrivals of resource type $i$ up to and inclusive of time $t$.

We will use $\bar{\lambdab}$ to denote a \textit{representative} distribution for the sequence $\{\lambdab^t\}_{t\in [T]}$. The relationship between $\bar{\lambdab}$ and $\{\lambdab^t\}_{t\in [T]}$ will be clear soon when we mention the General Position Gap condition we require on the sequence $\{\lambdab^t\}_{t\in[T]}$. For now, the reader may think of $\bar{\lambdab}$ as the mean of $\{\lambdab^t\}_{t\in[T]}.$
The model of the arrival process given so far is fairly general, and we need to add some constraints on the process so that it is compatible with the notion of offline resources (whose total inventory is known at time $t=0$). Towards this end, we will require that for $i \in \Ical^\off$, $\bar{\lambda}_i$ is deterministic, and $A^T_i = \bar{\lambda}_i \cdot T$ with probability 1. 
}

{\bf (Smoothed) General Position Gap assumption:}  We begin by formally defining the General Position Gap condition. Recall the static planning linear program \eqref{mp:SPP}.

\begin{definition}[General Position Gap]
\label{def:GPG}
{\blue Let $\xb^*$ be an optimal basic feasible solution to \SPP(${\lambdab}$), and let $\Mcal_+ = \{ m \in \Mcal  | x^*_m > 0\}$ denote the basic activities under this optimal solution. We say that $\lambdab$ satisfies the $\GPG_{\epsilon}$ condition if the following is true. \begin{enumerate}
    \item $\xb^*$ is the unique optimal solution with basis $\Mcal_+$, and,
    \item for any $\hat{\lambdab} \in \Bcal_{\epsilon,\TV} (\lambdab)$, there exists a unique optimal solution to \SPP($\hat{\lambdab}$) with $\Mcal_+$ as the basic activities.
    \end{enumerate}
    }
\end{definition}
The GPG condition stated in \cite{kerimov2021dynamic} requires that $\xb^*$  be unique and $x^*_m \geq \epsilon$ for all $m\in \Mcal_+$. Our condition is weaker in the sense that we allow non-unique optimal basis. If the optimal solution is unique (equivalently, there is a unique optimal basis) then the two definitions are equivalent (with a different choice of $\epsilon$). In the rest of the paper, we fix an optimal basic feasible solution $\xb^*$ and its basis $\Mcal_+$.{\blue \footnote{{\blue An example where the solution of the primal \eqref{mp:SPP} is non-unique is the following. Consider stochastic bin packing instance with bins of size $B=10$, and items of sizes 2,6 and 7. In particular $\lambda_6 = \lambda_7=0.4$ and $\lambda_2=0.2$ denote the probabilities with which an arrival is of size 6,7 and 2, respectively. In this case there are multiple optimal basis, for example $\{(2,2,6),(6),(7)\}$, $\{(2,6),(6),(7)\}$, $\{(6),(2,7),(7)\}$, where, e.g., $(2,2,6)$ denotes the configuration with two size 2 items and one size 6 item. However, the optimal solution to \eqref{mp:DSPP} is unique: $\alpha_2^*=0,\alpha_6^*=-1,\alpha_7^*=-1$. }
} }

The next assumption says that the $\GPG_{\epsilon_0}$ condition holds for the \textit{representative} distribution $\bar{\lambdab}$ with probability 1.
\begin{assumption}
\label{assum:GPGbar}
There exists an $\epsilon_0 > 0$, such that almost surely $\bar{\lambdab}$ satisfies $\GPG_{\epsilon_0}$. 
\end{assumption}

While Definition~\ref{def:GPG} is more transparent for the analysis in the current work, an equivalent, and perhaps easier to check definition can be arrived at via the Dual of \eqref{mp:SPP}.  
\begin{align}
\medskip
    \tag{\DSPP($\lambdab$)}
    \label{mp:DSPP}
    \min_{\{\alpha_i\}_{i\in \Ical}} &:  \sum_{i \in \Ical} \lambda_i \cdot \alpha_i  \\
    \nonumber
    \mbox{subject to} &  \\
        \nonumber
    \forall m \in \Mcal  &:   \sum_{i \in \Ical} M_{im} \cdot \alpha_i \geq r_m. 
\end{align}

{\blue The next proposition says that the General Position Gap condition is equivalent to the uniqueness of the optimal solution to \eqref{mp:DSPP}.

\begin{proposition}\label{prop:GPG_dual}
The distribution $\lambdab$ satisfies $\GPG_{\epsilon}$ for some $\epsilon>0$ if and only if for all $\hat{\lambdab}\in \Bcal_{\epsilon,\TV}(\lambdab)$, $\alphab^*$ is the optimal solution to $\DSPP(\hat{\lambdab})$. That is, the optimal solution $\alphab^*$ of \eqref{mp:DSPP} is unique.
\end{proposition}
The $\eqref{mp:DSPP}$ problem can equivalently be written as minimizing $\alphab^\top \lambdab$ subject to $\alphab \in \Acal$ where $\Acal$ is the polyhedron given by the inequalities in $\eqref{mp:DSPP}$ and which is independent of $\lambdab$. For almost all $\lambdab$, except for a measure 0 set, this minima is attained uniquely at a corner of $\Acal$. }

Finally, we state the assumption we need on the stochastic process $\{\lambdab^t\}_{t \in [T]}$.

\begin{assumption}[Smoothed $\GPG_{\delta, \tau}$]
\label{assum:smoothed} 
There exist a constant $\delta < \epsilon_0$, a smoothing window size $\tau \in \NN$, {\blue and a $\Fcal^t$-predictable stochastic process $\{\hat{\lambdab}^t\}_{t \in [T]}$, which we call \textit{smoothed distributions},} such that
\begin{enumerate}
\item $\hat{\lambdab}^t \in \Bcal_{\epsilon_0-\delta,\TV}(\bar{\lambdab})$ for each $t \in [T]$, and
\item $ \sum_{s=1}^{t-2\tau} \hat{\lambdab}^s \leq  \sum_{s=\tau+1}^t \lambdab^s \leq \sum_{s=1}^{t} \hat{\lambdab}^s$, for all $t\in[\tau+1,T]$.
\end{enumerate}
\end{assumption}

The smoothed $\GPG_{\delta,\tau}$ condition relaxes the requirement that the item distribution be \textit{i.i.d.} as well as that the distribution at each time step be close enough to $\bar{\lambdab}$. We need this flexibility to incorporate offline resources, since our emulation Algorithm~\ref{alg:emulation} injects offline resources on a deterministic schedule. Instead, we require that the \textit{smoothed} distributions $\hat{\lambdab}^t$ be close enough to $\bar{\lambdab}$, and that their cumulative sum tracks the cumulative sum of $\lambdab^t$ with bounded error. {\blue Since the smoothed $\GPG_{\delta, \tau}$ assumption and the arrival process model are stated quite abstractly we give a few examples where these hold:
\begin{enumerate}
\item Let the sequence be $\{\lambdab^t\}$ is deterministic, and each $\lambdab^t \in \Bcal_{\epsilon_0-\delta,\TV}(\bar{\lambdab})$ for some $\bar{\lambdab}$ satisfying Assumption~\ref{assum:GPGbar}. In this case $\GPG_{\delta, \tau}$ condition holds with $\tau=0$.
\item Let the online resources arrive \textit{i.i.d.}, and the offline items `arrive' via the construction in Section~\ref{sec:emulation}. Then either $\lambdab^t = \lambdab^\on$, or $\lambdab^t=\eb_i$ for some $i\in \Ical^
\off$. Lemma~\ref{lem:bounded_discrepancy} shows that the smoothed $\GPG_{\delta,\tau}$ condition is true with $\delta = \epsilon_0$ (i.e., $\hat{\lambdab}^t = \bar{\lambdab}$) 
and $\tau = (n^\off+1) / \bar{\lambda}_{\min}$.
\item Let the sequence of online resources be sampled \textit{i.i.d.} from $\bar{\lambdab}$, and then the sample path  be ``locally permuted'' by an adversary before the resources arrive at the matching system. By locally permute, we mean that there exists an integer $\gamma$ so that no item is moved more than $\gamma$ positions from its initial position. Such a scenario can happen for example if the items pass through a processing stage that is modeled as a queueing system before arriving at the matching system, or with batch arrivals. While the process $\lambdab^t$ might be fairly complex to model explicitly, the $\GPG_{\delta,\tau}$ condition still holds for this case with $\delta = \epsilon_0$ (i.e., $\hat{\lambdab}^t=\bar{\lambdab}$) and $\tau = \gamma$.
\end{enumerate}
}

\vspace{3mm}

{\bf Matching process:} Let $X^t_m$ denote the number of matches of configuration $m$ completed up to and inclusive of time $t$. Define:
\[ N^t_i := A^t_i -\sum_{m \in \Mcal} X^{t}_m \cdot M_{im} \]
as the units of resource $i$ which have arrived but are unmatched up to time $t$.

For a configuration $m$ which involves only offline and online-queueable resource types, a match can be completed at any time there are enough resources available. That is, if $ N^{t-1}_i + Z^t_i \geq M_{im} $ for all $i \in \Ical^\onq$, and $ \bar{\lambda}_i \cdot T - \sum_{m'} X^{t-1}_{m'} \cdot M_{im'} \geq M_{im} $ for all $i \in \Ical^\off$. Note that even though we model the offline resources as arriving online, this is only for bookkeeping for our Sum-of-Squares algorithm; the feasibility of a match depends on the total availability. That is, it is possible that $N^t_i < 0$ for $i \in \Ical^\off$.

For a configuration $m$ which involves an (exactly one) online-nonqueueable resource unit, a match can be completed only at the time of arrival of this resource as long as there are enough resources available. This criterion is the same as above: $N^{t-1}_i + Z^t_i \geq M_{im}$ for all $i \in \Ical^\onq$, and $ \bar{\lambda}_i \cdot T - \sum_{m'} X^{t-1}_{m'} \cdot M_{im'} \geq M_{im} $ for all $i \in \Ical^\off$. 

\vspace{3mm}

\vspace{3mm}

{\bf Main Result:} Our first main result is Theorem~\ref{thm:mainUB} which proves that the sum-of-squares algorithm described in Section~\ref{sec:SoS} gets bounded anytime regret for online multiway matching if no configurations contain both online-queueable and an online-nonqueueable resource, and $\Ocal(\log t)$ anytime regret otherwise. The second result, Theorem~\ref{thm:mainLB}, complements the above by proving that in the latter case $\Omega(\log t)$ anytime regret is unavoidable.

\begin{theorem}
\label{thm:mainUB}
Under the non-stationary arrival process given by the sequence ${\lambdab}^t$ satisfying $\GPG_{\epsilon_0}$ and the smoothed $\GPG_{\delta, \tau}$ conditions for constants $0 < \delta < \epsilon_0$, and some smoothing window bound $\tau$, the Sum-of-Squares algorithm (Algorithm~\ref{alg:SoS}) satisfies the following regret guarantees:
\begin{itemize}
\item If there are no configurations in $\Mcal_+$ which contain resources from both $\Ical^\onq$ and $\Ical^\onnq$, then for any time $t \in [T]$ 
\[ \expct{ \sum_{m} r_m \cdot X^t_m }  \geq \expct{\OPT(\Ab^t)} - \frac{\kappa_1(n,\tau)}{\delta}, \]
where $\kappa_1(n, \tau)$ is a constant that is polynomial in $n,\tau$, independent of $T$.
\item If there are feasible configurations in $\Mcal_+$ which contain resources from both $\Ical^\onq$ and $\Ical^\onnq$, then for any time $t \in [T]$ 
\[ \expct{ \sum_{m} r_m \cdot X^t_m }  \geq \expct{\OPT(\Ab^t)} - \frac{\kappa_2(n,\tau)}{\delta} \log t, \]
where $\kappa_2(n, \tau)$ is a constant that is polynomial in $n,\tau$, independent of $T$.
\end{itemize}
\end{theorem}
As mentioned earlier, in the case where online resources arrive \textit{i.i.d.}, the smoothed $\GPG_{\delta,\tau}$ condition is true with $\delta = \epsilon_0$ and $\tau = (n^\off+1) / \lambda_{\min}$, and therefore the regret for the first case in Theorem~\ref{thm:mainUB} as a function of the GPG parameter $\epsilon_0$ grows as $1/\epsilon_0$. This growth rate was shown to be necessary {\blue when all resources are only of online-queueable type} in \cite{kerimov2021dynamic}.

The next theorem shows that a regret of $\Ocal(\log t)$ is unavoidable in general when configurations involve both online-queueable and online-nonqueueable resources.

\begin{theorem}
\label{thm:mainLB}
There is a multiway matching instance with online-queueable and online-nonqueueable resources where no online algorithm can guarantee for all times $t \in [T]$, and for a small enough constant $c>0$:
\[ \expct{ \sum_{m} r_m \cdot X^t_m }  \geq \expct{\OPT(\Ab^t)} -  c \log t. \]

\end{theorem}

\section{Sum-of-Squares Algorithm}
\label{sec:SoS}

Our proposed algorithm exploits the knowledge of the optimal basic activities for $\SPP(\bar{\lambdab})$ by only ever creating matching configurations from the set $\Mcal_+$. When a resource of type $i$ arrives at time $t$, it is irrevocably committed to a basic configuration $m \in \Mcal_+$ based on a Sum-of-Squares potential associated with each configuration that we define shortly. For each configuration $m$ and resource $i \in \Ical(m)$ we define two queue lengths -- virtual queue lengths, which are for book-keeping and used to define the sum-of-squares potential for configuration $m$; and true queue lengths which correspond to the resources that have been committed to $m$ thus far but not departed the system due to a match. We also define two kinds of matches -- virtual matches, which are for book-keeping and necessitated by the presence of online-nonqueueable resources; and true matches which are the physically realized matches.

We now describe the Sum-of-Squares potential function and the algorithm more formally.

\vspace{3mm}

{\bf Virtual/true matches and queue lengths:} We denote by $Y^t_{im}$ the number of resource units of type $i$ committed by our algorithm to matching configuration $m \in \Mcal_+$ up to and including time $t$. Recall that $X^t_m$ denotes the number of matches completed up to time $t$ for configuration $m$. In addition to $X^t_m$, which we will call \textit{true} matches, we define an analogous quantity $\Xs^t_m$ for the number of \textit{virtual matches} completed up to time $t$ for configuration $m$. The notion of virtual match is primarily used for configurations with items in $\Ical^\onnq$, where we execute a virtual match for the purposes of bookkeeping every time the item from $\Ical^\onnq$ is committed to the configuration, even if enough resources are not available to complete a match. Only a subset of virtual matches are true matches -- when enough resources are available. Based on the above definitions, we define the \textit{true queue length} of type $i$ resources for configuration $m$ by:
\begin{align}
\label{eqn:virtualN}
    Q^t_{im} &= Y^t_{im} - X^t_{m} \cdot M_{im},
\end{align}
and \textit{virtual queue length} of type $i$ resources for configuration $m$ by:
\begin{align}
\label{eqn:trueQ}
    \Qs^t_{im} &= Y^t_{im} - \Xs^t_{m} \cdot M_{im}.
\end{align}
{\blue ($\Qs$ should be read as script-$Q$; similarly $\Xs$ should be read as script-$X$. The scripted symbols are used for virtual versions of the true quantities).}
We always have $X^t_m \leq \Xs^t_m$ and therefore $Q^t_{im} \geq \Qs^t_{im}$. The virtual queue lengths for items in $\Ical^{\off}\cup \Ical^{\onq}$ and configurations with a resource in $\Ical^\onnq$ can become negative -- this is a feature since negative virtual queues incentivize our algorithm to commit more resources to such configurations. 

\vspace{3mm}

{\bf Sum-of-Squares potential function:} For a configuration $m$, with virtual queue length vector $\Qsb_m = (\Qs_{im})_{i \in \Ical}$, we define the Sum-of-Squares potential as follows:
\begin{align}
\label{eqn:SS}
    \Phi_m(\Qsb_m) = \left(\Qs_{\sigma_m(1) m} - \Qs_{\sigma_m(\ell_m)m}  \right)^2 + \sum_{k=2}^{\ell_m} \left(\Qs_{\sigma_m(k)m} - \Qs_{\sigma_m(k-1)m}\right)^2, 
\end{align} 
where $\sigma_m$ denotes an arbitrary permutation of $\Ical(m)$ that is fixed throughout the execution of the algorithm. Intuitively, minimizing the Sum-of-Squares potential equalizes the virtual queue lengths involved in configuration $m$. For every configuration one of the virtual queue lengths is always 0. In the case of a configuration with a resource in $\Ical^\onnq$ the virtual queue length of this resource is 0. For the other configurations, some virtual queue length must be 0 otherwise a virtual match would be completed until this was the case. This fact, combined with the spirit of trying to equalize queue lengths through the potential function \eqref{eqn:SS}, achieves the goal of keeping all queues small. This is also precisely the intuition behind the back-pressure algorithm of \cite{tassiulas1990stability} for routing packets/flows in networks. However, unlike network flow, in our case there is no natural ordering of virtual queues, and therefore we use the cyclic Sum-of-Squares potential function.

As a quick remark, in the case where $M_{im}$ take general values (other than $0$ or $1$), the Sum-of-Squares potential for configuration would be defined as:
\begin{align}
\label{eqn:SS}
    \Phi_m(\Qsb_m) = \left(\frac{\Qs_{\sigma_m(1) m}}{M_{\sigma_m(1) m}} - \frac{\Qs_{\sigma_m(\ell_m)m}}{M_{\sigma_m(\ell_m)m}}  \right)^2 + \sum_{k=2}^{\ell_m} \left( \frac{\Qs_{\sigma_m(k)m}}{M_{\sigma_m(k)m}} - \frac{\Qs_{\sigma_m(k-1)m}}{M_{\sigma_m(k-1)m}}  \right)^2.
\end{align}

We denote the overall Sum-of-Squares potential as:
\begin{align}
    \Phi(\Qsb) &= \sum_{m \in \Mcal_+} \Phi_m(\Qsb_m).
\end{align}

The change in potential function when an item of type $i$ is potentially committed to $m$ is computed as follows:
\begin{align}
    \label{eqn:deltaSS}
    \Delta \Phi_m(\Qsb_m,i) &= \Phi_m(\Qsb'_m) - \Phi_m(\Qsb_m)
    \intertext{where $\Qsb'_m$ is defined as}
    \nonumber
    \Qsb'_m &= 
        \begin{cases}
        \Qsb_m + \eb_i - \Mb_m  & \mbox{if \textsf{virtual\_feasible}($m,\Qsb_m,i$)},\\
        \Qsb_m + \eb_i  & \mbox{otherwise}.
        \end{cases}
        \intertext{The boolean function \textsf{virtual\_feasible}($m,\Qsb_m,i$) is given by:}
\label{eqn:virtual_feasible}
    \textsf{virtual\_feasible}(m,\Qsb_m,i) &= 
        \begin{cases}
            \true & \mbox{if $i \in \Ical^\onnq$,} \\
            \true & \mbox{if $\Ical^\onnq \cap \Ical(m) = \emptyset$ and $\Qsb_m+\eb_i \geq \Mb_m$,} \\
            \false & \mbox{otherwise.}
        \end{cases}
\end{align}
In words, we always execute a virtual match whenever an online-nonqueueable resource is committed to the configuration. For configurations without any online-nonqueueable resource, we only execute a virtual match when there are sufficient number of true resources available in the queues. Note that even in this case, a virtual match does not always imply a true match since the resources in $\Ical^\off$ could have been exhausted.

Analogous to \textsf{virtual\_feasible}, we define the boolean function \textsf{true\_feasible}:
\begin{align}
\label{eqn:true_feasible}
    \textsf{true\_feasible}(m,\Qb_m,\Xb,i) &= 
        \begin{cases}
            \false & \mbox{if $\exists i' \in \Ical^\off \cap \Ical(m) : \bar{\lambda}_{i'} T < M_{i'm} + \sum_{m'}X_{m'} M_{i'm'}$,} \\
            \false & \mbox{if $\exists i' \in \Ical^\on \cap \Ical(m) : Q_{i'm}+\ind\{i'=i\} < M_{im}$,} \\
            \true & \mbox{otherwise.}
        \end{cases}
\end{align}
In words, a true match is \textit{infeasible}, if either the total number of offline resources is insufficient (hence the dependence on $\Xb$, the number of true matches executed so far), or the true queue length for any online resource in $\Ical(m)$ is insufficient; otherwise the true match is feasible. In our algorithm, we check the feasibility of a true match only if a virtual match is executed.

Algorithm~\ref{alg:SoS} is our proposed greedy Sum-of-Squares ($\SoS$) algorithm.
\begin{algorithm}[ht]
\small
   \caption{Sum-of-Squares algorithm $(\SoS)$ for Multiway matching}
   \label{alg:SoS}
\DontPrintSemicolon
 \SetKw{KwDef}{Definition:} 
 \SetKw{KwInit}{Initialize:} 
 \SetAlgoLined
 \KwInit{} $X^0_m = \Xs^0_m = 0$ for each $m \in \Mcal_+$\\ \hspace{0.4in}  $Y^0_{im}=0$ for all $m \in  \Mcal_+, i \in \Ical(m)$ 
 
 \KwDef{} $Q^t_{im}, \Qs^t_{im}$ in \eqref{eqn:virtualN}-\eqref{eqn:trueQ};  $\Delta \Phi_m$ in \eqref{eqn:deltaSS}; \textsf{virtual\_feasible}, \textsf{true\_feasible} in \eqref{eqn:virtual_feasible}-\eqref{eqn:true_feasible}
 
 \For{$t=1, \ldots, T$}{
     Observe arrival type $I^t$
     
     Let $m^t = \argmin_{m} \left\{  \Delta \Phi_m(\Qsb^{t-1}_m, I^t)  \ | \ m \in \Mcal_+, I^t \in \Ical(m) \right\}$ \tcp*[r]{Find configuration to commit}
     
     $Y^t_{I^t m^t} = Y^{t-1}_{I^t m^t} +1 $ \tcp*[r]{Increment resource-configuration counter}
     
     \If{
        {\upshape \textsf{virtual\_feasible}$(m^t,\Qsb^{t-1}_{m^t},I^t)$}
     }
     {
     $\Xs^t_{m^t} = \Xs^{t-1}_{m^t}+1$ \tcp*[r]{Increment virtual match counter}
     
        \If{\upshape \textsf{true\_feasible}$(m^t,\Qb^{t-1}_{m^t},\Xb^{t-1},I^t)$}{
         $X^t_{m^t} = X^{t-1}_{m^t}+1$ \tcp*[r]{Increment true match counter}
        }
     }
     \For{$m \neq m^t$}{
         $X^t_m = X^{t-1}_m ; \ Y^t_m = Y^{t-1}_m$
         
     }
     \For{$(i,m) \neq (I^t,m^t)$}{
        $X^t_{im} = X^{t-1}_{im}$
        
     }
     }
\end{algorithm}

 {\blue 
 \begin{remark} If we had used the true queue length $Q_{i'm}$ as the criterion for offline resources in $\textsf{true\_feasible}$, instead of the global availability, then our algorithm would get a regret of $\Ocal(\log t)$ for configurations with online-nonqueueable and offline resources. In particular, we would not get $\Ocal(1)$ regret for Network Revenue Management application.
 \end{remark}}
{\blue
\begin{remark} Contrast with use of virtual queues in \cite{nazari2019reward}:  In \citep{nazari2019reward}, the authors look at multiway matching with only online-queueable items, and also use a notion of virtual queues. However, their use differs from the use of virtual queues in this paper. In the algorithm of \cite{nazari2019reward}, there is one true queue and one virtual queue per item type. Both increase on item arrivals. However, virtual queues decrease whenever virtual matches are executed, and the identity of the virtual match configuration is decided based on optimizing a combination of the revenue $r_m$ and the current virtual queue lengths. A virtual match is added to a list of incomplete matches which are converted into true matches when there are enough resources available in the true queues. There are two differences in our paper: (i) there is a virtual queue for each item-configuration pair, and (ii) it is the assignment of items to match configurations that happens based on virtual queues; virtual/true matches execute whenever there are enough virtual/true resources committed to a match configuration. The use of virtual queues in \citep{nazari2019reward} allows them to adapt to unknown arrival rates $\lambda$ without explicitly learning them, as the virtual queue lengths converge to the dual variables of the $\SPP$. However, their algorithm requires tuning a parameter $\beta$, roughly corresponding to the inverse of the virtual queue lengths, to be ``as large as possible, but not larger.'' 
\end{remark}
}

\section{Regret Analysis}
\label{sec:analysis}

In Section~\ref{sec:regret_upper} we prove Theorem~\ref{thm:mainUB} which upper bounds the any-time regret for Algorithm~\ref{alg:SoS}. In Section~\ref{sec:regret_lower} we prove Theorem~\ref{thm:mainLB} which lower bounds the anytime regret for any online algorithm in the case of configurations with both $\Ical^\onq$ and $\Ical^\onnq$ resources by $\Omega(\log t)$.

\subsection{Regret upper bound for Sum-of-Squares algorithm}
\label{sec:regret_upper}
We prove the regret bound in three steps. 

{\bf Step 1 (Bounding the gap between true reward and virtual reward):} Note that for any configuration $m$ and time $t$, the virtual reward $r_m \Xs_m^t$ is an upper bound on the true reward $r_m X_m^t$. Our first lemma bounds this gap. 
\begin{lemma} 
\label{lem:gap_virtual_true}
For a configuration $m$ and time $t$, the gap between online and offline matches is bounded as follows:\\
\underline{\textit{Case 1:}} $m$ with both online-queueable and online-nonqueueable resources ($\Ical(m) \cap \Ical^\onq\neq \emptyset$, and $\Ical(m) \cap \Ical^\onnq \neq \emptyset$):
\[ \Xs_m^t - X_m^t \leq  \max_{ i \in \Ical(m) \cap \Ical^\onq} \max_{s \in [t] } \left(\Qs_{im}^s  \right)^- + \sum_{ i \in \Ical(m) \cap \Ical^\off } \  \sum_{ m' : i \in \Ical(m')} \left(\Qs_{im'}^t\right)^- + |\Ical(m)|.  \]
\underline{\textit{Case 2:}} Otherwise; $m$ without online-nonqueueable resource, or without online-queueable resource:
\[ \Xs_m^t - X_m^t \leq  \sum_{ i \in \Ical(m) \cap \Ical^\off } \  \sum_{ m' : i \in \Ical(m')} \left(\Qs_{im'}^t\right)^-  + |\Ical(m)|.  \]    
Case 2 implies that for configurations with only  $\Ical^\onq$ resources, the number of virtual and true matches are the same.
\end{lemma}
Lemma~\ref{lem:gap_virtual_true} was not required for the case studied in Section~\ref{sec:iid} since we did not need virtual queues in that setting.
\vspace{3mm}

{\bf Step 2 (Bounding the gap between virtual reward and offline optimal  reward):} The next sample path lemma states that if $\Ab^t/t$ is close to $\bar{\lambdab}$, then the size of the virtual queues give a bound on the gap between the offline optimal reward and the virtual reward.
\begin{lemma}
\label{lem:gap_opt_virtual}
    If $\Ab^t/t \in \Bcal_{\epsilon_0,\TV}(\bar{\lambdab})$, then, 
    \[ \OPT(\Ab^t) - \sum_m r_m \cdot \Xs_m^t =  \rb_{\Mcal_+}^\top \Mb_{\Mcal_+}^{-1} \sum_{m \in \Mcal_+} \Qsb^t_m. \]
    In the above, $\rb_{\Mcal_+}$ denotes the vector $(r_m)_{m\in \Mcal_+}$, and $\Mb_{\Mcal_+}$ denotes the matrix $\left( \Mb_{m} \right)_{m \in \Mcal_+}$. 
\end{lemma}

{\blue As we point out in the proof of Proposition~\ref{prop:GPG_dual}, $\rb_{\Mcal_+}^\top \Mb_{\Mcal_+}^{-1}$ is the optimal solution to the dual \eqref{mp:DSPP}, and can be thought of as the value of resources. Therefore, the right hand side of the expression in Lemma~\ref{lem:gap_opt_virtual} can be thought of as the dual price of unused resources in the virtual queues.} As a corollary of this sample path lemma, we get the following bound on the gap between the expected offline optimal reward and the expected virtual reward, which generalizes Lemma~\ref{lem:regret_decomposition} to non-stationary arrival instances.

\begin{lemma}
\label{lem:bounded_regret}
    If $\{{\lambdab}^t\}$ satisfies the smoothed $\GPG_{\delta,\tau}$ condition for $\delta > 0$, then, 
    \[ \expct{ \OPT(\Ab^t) } - \expct{\sum_m r_m \cdot \Xs_m^t} \leq  \rb_{\Mcal_+}^\top \Mb_{\Mcal_+}^{-1} \sum_{m \in \Mcal_+} \expct{\Qsb^t_m} + \Ocal(\tau). \]
\end{lemma}

\vspace{3mm}

{\bf Step 3 (Bounding $\Qs_{im}^t$ and $\min_{s \in [t]} \Qs_{im}^s$):} To apply Lemma~\ref{lem:gap_virtual_true} we need to show that the virtual queues remain small (either in expectation for any fixed time $t$, or the minimum over the interval $[t]$). As a first step towards this, in the next lemma we show that a carefully constructed Lyapunov function that upper bounds the sum-of-squares potential $\Phi(\Qsb^t)$ exhibits a useful negative drift condition. Proving this Lemma is the main technical novelty of the paper (compared to the drift analysis in Section~\ref{sec:iid} which followed in straightforward manner from \citep{csirik2006sum}).  

\begin{lemma}
\label{lem:drift_lemma}
There exists a non-negative stochastic process $\Xi^t$, and a potential function $\Psi^t = \sqrt{\Phi(\Qsb^t) + \Xi^t }$ for Algorithm~\ref{alg:SoS}, satisfying, for all $t \in [T]$,
\[ 0 \leq \Xi^t \leq 2 \tau \left( \sqrt{ 8 \Phi(\Qsb^t) }+ 8\tau + 2 \ln (1+4\tau) \right), \]
and such that $\Psi^t$ obeys the following Lyapunov-drift conditions for $t \geq \tau + 1$ under the smoothed $\GPG_{\delta, \tau}$ assumption:
\begin{enumerate}[label=(\alph*)]
\item Bounded variation:  $\left| \Psi^{t+1} - \Psi^t \right| \leq K_1(\tau)$
\item Expected Decrease:  
\[ \expct{\Psi^{t+1} - \Psi^t | \Psi^t \geq K_2(\tau, \delta))} \leq - \frac{\delta}{8\sqrt{2}n^2}. \]
\end{enumerate}
In the above,
\begin{align*}
    K_1(\tau) &:= 6 \tau^2 + 9 \tau + 2\tau \log(3+4\tau) + 8, \\
    K_2(\tau,\delta) & := \sqrt{2\max\{c_1(\tau, \delta), c_2(\tau)\}} = \Ocal\left( n^2\tau^2/\delta \right).
\end{align*}
where $c_1(\tau,\delta), c_2(\tau)$ are constants defined in \eqref{eqn:c1} and \eqref{eqn:c2}, respectively.
\end{lemma}

A result of \cite{hajek1982hitting} shows that under the conditions of Lemma~\ref{lem:drift_lemma}, for some constant $\eta > 0$, $\expct{e^{\eta \Psi^t}}$ is bounded for all $t$, as well as that this process is a super-martingale in the sense  $\expct{ e^{\eta \Psi^{t+1}} | \Psi^t \geq K_2(\tau, \delta)}\leq \rho e^{\eta \Psi^t}$ for some $0\leq \rho < 1$. However a direct application of the bounds from \cite{hajek1982hitting} give us an expected regret of $\Ocal(1/\delta^2)$, not $\Ocal(1/\delta)$. Instead we will use the bound we proved in Lemma~\ref{lem:icx_bound} to upper bound the process $\Psi^t$ in increasing convex order by a simple reflected random walk to finish the proof.

\begin{proofof}{Theorem~\ref{thm:mainUB}}
We apply Lemma~\ref{lem:icx_bound} by choosing $\Upsilon^t := \Psi^t, K_1 := K_1(\tau), K_2 := K_2(\tau,\delta), \delta' := \frac{\delta}{16\sqrt{2}n^2 K_1(\tau)} $ to conclude that $\left\{ \Psi^t\right\}_{t\in [\tau,T]}  \leqicx \left\{ \Gamma^t\right\}_{t\in [\tau,T]}$.
Choosing $f(\Xb) = X^t$ in Definition~\ref{defn:icx}, and using Lemma~\ref{lem:icx_bound}, we get $\expct{\Psi^t} \leq \expct{\Gamma^t}$ for all $t \geq \tau$. Defining $\rho = \frac{1-2\delta'}{1+2\delta'}$, a straightforward calculation shows that the expectation of the steady state distribution for the random walk $\{\Gamma^t\}$ is $\expct{\Gamma^\infty} = \Gamma_{\min}+K_1 \frac{\rho}{1-\rho}$. Since the random walk $\{\Gamma^t\}$ starts at $\Gamma^\tau$, and not in steady state, a simple coupling argument shows that we can bound
\[ \expct{\Gamma^t} \leq \max\{\Gamma^\tau, \Gamma_{\min}\} + K_1 \frac{\rho}{1-\rho} = \Ocal( n^2\tau^4/\delta). \]
Since for each $m\in \Mcal_+$, $i \in \Ical(m)$, $\expct{\Qs^t_{im}} \leq \expct{\Phi^t} \leq  \expct{\Psi^t}$, the above implies that the expected length of virtual queues is bounded by a constant depending on ($n, \tau, \delta$), and combined with Lemma~\ref{lem:gap_virtual_true} (Case 2) and Lemma~\ref{lem:gap_opt_virtual} gives bounded anytime regret for multiway matching when no configuration has both online-queueable and online-nonqueueable resources.

We next proceed to prove logarithmic regret for the case when some configurations have both online-queueable and online-nonqueueable resources, in which case according to Case 1 of Lemma~\ref{lem:gap_virtual_true}, we need to bound $\expct{\max_{s \in [t]} \left( \Qs^s_{im} \right)^-}$, which is in turn bounded by $\expct{\max_{s \in [t]} \Psi^s}$. Since $\max$ is an increasing convex function, we further have the bound, $\expct{\max_{s \in [t]} \Psi^s} \leq \expct{\max_{s \in [t]} \Gamma^s}$.

To bound $\expct{\max_{s \in [t]} \Gamma^s}$, consider a random walk on $\NN$ with initial state $1$, reflection boundary $0$, and probabilities of step size of $+1, -1$ being $\frac{1}{2}-\delta'$ and $\frac{1}{2}+\delta'$, respectively. Let $S$ denote the random variable for the supremum of this random walk until the process first hits $0$. We can write the following recurrence inequality for the distribution of $S$:
\begin{align*}
    \prob{S \geq 1} &= 1, \\
    \prob{S \geq 2} &= \left( \frac{1}{2} -\delta'\right), \\
\mbox{for $x\geq 3$},\     \prob{S \geq x} & \leq  \left( \frac{1}{2} + \delta'\right) \cdot 0 + \left( \frac{1}{2} - \delta'\right) \left( \prob{S \geq x-1} + \prob{S \geq x}\right)\\
& \leq \rho \prob{S \geq x-1} \leq \left( \frac{1}{2} -\delta'\right) \rho^{x-2}.
\end{align*}

In words, the supremum of $\Gamma^{t}$ in one excursion above $\Gamma_{\min}$ until it hits this level again is stochastically bounded from above by $3 K_1 $ plus a geometrically distributed random variable with mean $\frac{K_1}{1-\rho}\leq \frac{K_1}{\delta'} $. Since there can be at most $t$ such excursions in the interval $\{1,\ldots, t\}$, and the expectation of the maximum of $t$ i.i.d. geometrically distributed random variables, each with mean $K_1/\delta'$, is given by $H_t\frac{K_1}{\delta'}$ ($H_t \sim \log t $ is the $t$th harmonic number), we get,
\[ \expct{\max_{s \in [t]} \Psi^s } \leq \expct{\max_{s \in [t]} \Gamma^s } \leq  \Gamma_{\min} + 3 K_1 + \frac{K_1}{\delta'} H_t = K_2(\tau, \delta) + \Ocal\left(  \frac{ n^2 \tau^4 \log t}{\delta} \right).\]
Combined with Lemma~\ref{lem:gap_opt_virtual} (Case 1) and Lemma~\ref{lem:gap_virtual_true}, the above proves the second part of Theorem~\ref{thm:mainUB}.
\end{proofof}

\subsection{Regret Lower bound}
\label{sec:regret_lower}

\begin{figure}[t]
    \centering
    \includegraphics[width=2.2in]{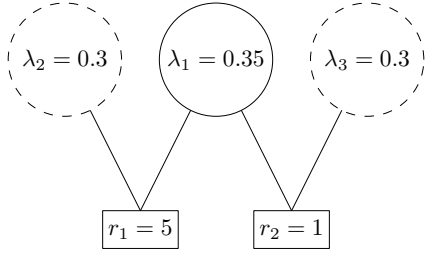}
    \caption{Instance used for proving Theorem~\ref{thm:mainLB}. Online-queueable resources correspond to solid circles, and online-nonqueueable to dashed circles. The matching configurations are represented by rectangles.}
    \label{fig:LB_instance}
\end{figure}

Our goal is to show that, in general, for matching instances where configurations involve both online-queueable and online-nonqueueable resources, there must be some time $t$ at which an expected regret of $\Omega(\log t)$ compared to the offline optimal for the instance seen until $t$ is unavoidable. To prove this, we consider one specific instance, illustrated in Figure~\ref{fig:LB_instance}. There are two matching configurations, $\Mcal = [2]$ and three resource types with $\Ical^{\onq}=\{1\}$ and $\Ical^{\onnq}=\{2,3\}$. The matching matrix is 
{\blue
\begin{align*}
\renewcommand*{\arraystretch}{1.5}
\Mb = \begin{bmatrix}
1 & \quad 1 \\
1 & \quad 0 \\
0 & \quad 1  \end{bmatrix}. 
\end{align*}
}
The solution of $\SPP(\lambdab)$ is $x_1 = 0.3, x_2=0.05$. 

In words, only resource 1 can be queued and both resource type 2 and 3 are only available for matching on arrival. Since matches with resource type 2 generate higher reward, and resource type 2 are scarcer than resource type 1, any time a resource of type 2 arrives and there are no resources of type 1 in the queue, we incur  additional regret. The key intuition behind the regret lower bound of $\Omega(\log t)$ is that if at some time $t$, the expected number of units of resource 1 in the queue is $\Omega(\log t)$, then we have already incurred a regret of $\Omega(\log t)$ \textit{at this time}, since the offline optimal restricted to arrivals in $[t]$ would have matched all resources of type 1. Instead, if we have an expected queue length of $o(\log t)$ at all times, then with high probability, there can be a sequence of $\Omega(\log T)$ consecutive arrivals of type 2, most of which are lost due to insufficient type 1 resources. This causes us to incur $\Omega(\log T)$ regret at the terminal time $T$.

\begin{lemma}
\label{lem:LB_regret}
For the instance shown in Figure~\ref{fig:LB_instance}, there exists a constant $c > 0$, and an integer $T_0$, such that for all $T \geq T_0$, for any online algorithm, either 
\begin{enumerate}
    \item there are exists some $t \in [T/2, T]$ such that $\expct{N^t_1} \geq 2c \log T $, and therefore,
    \[ \expct{r_1 X_1^t + r_2 X_2^t} \leq \expct{\OPT(\Ab^t)} - c \log T, \]
    \item or $\expct{X^T_1} \leq A^T_2 - c \log T$, and therefore,
    \[ \expct{r_1 X_1^T + r_2 X_2^T} \leq \expct{\OPT(\Ab^T)} - 2 c \log T. \]
\end{enumerate} 
\end{lemma}

\begin{proofof}{Theorem~\ref{thm:mainLB}}
The proof immediately follows from Lemma~\ref{lem:LB_regret}. If the first case of the lemma applies, then the regret at such a time is $\Omega(\log T)$, otherwise the second case applies and the regret at time $T$ is $\Omega(\log T)$.
\end{proofof}

\section{Experiments} 
\label{sec:experiments}

\begin{figure}[t]
    \centering
    \includegraphics[width=5in]{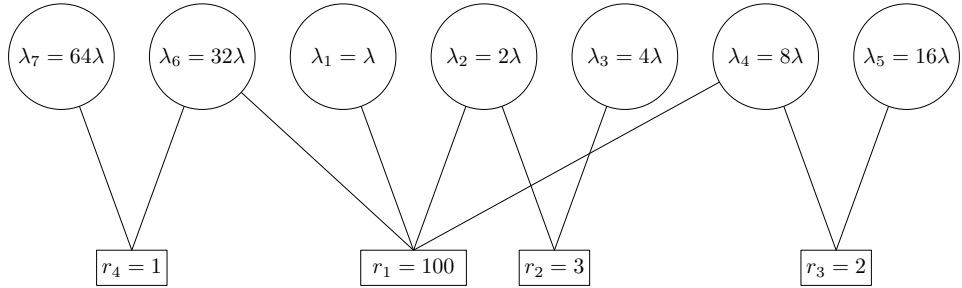}
    \caption{Multiway matching instance with only online-queueable resources (Figure 9 from \cite{kerimov2021dynamic}). $\lambda$ is the normalization constant.
    }
    \label{fig:multiway_instance}
\end{figure}

\begin{figure}[ht!]
\begin{center}
\subfigure{\includegraphics[width=3.2in]{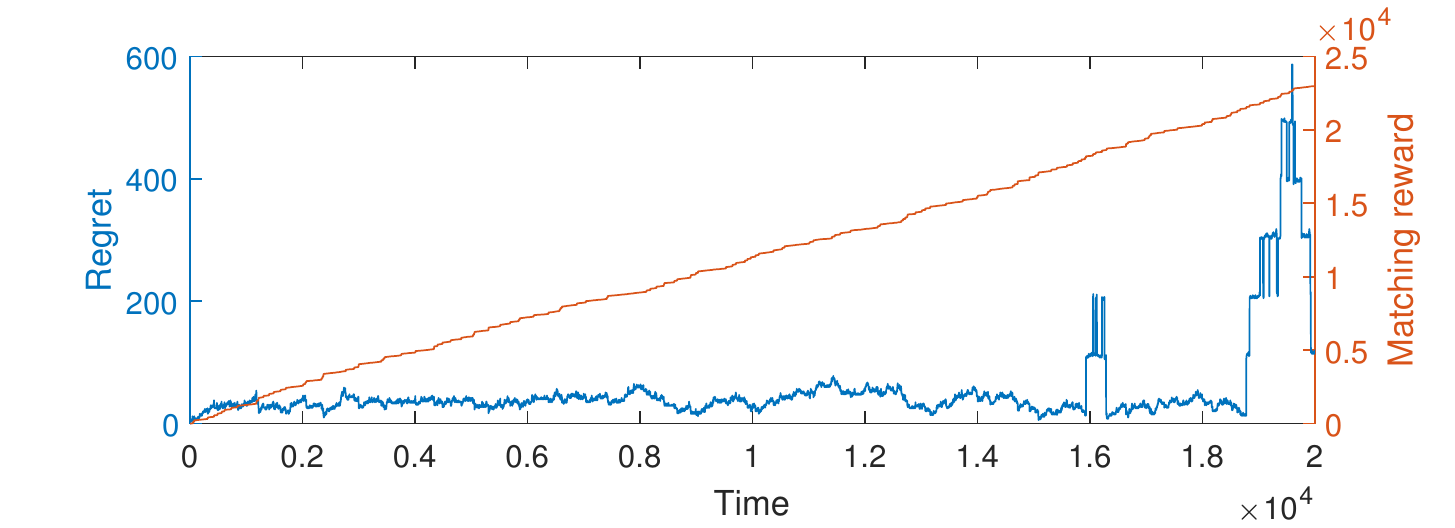}}
\hspace{-0.1in}
\subfigure{\includegraphics[width=3.2in]{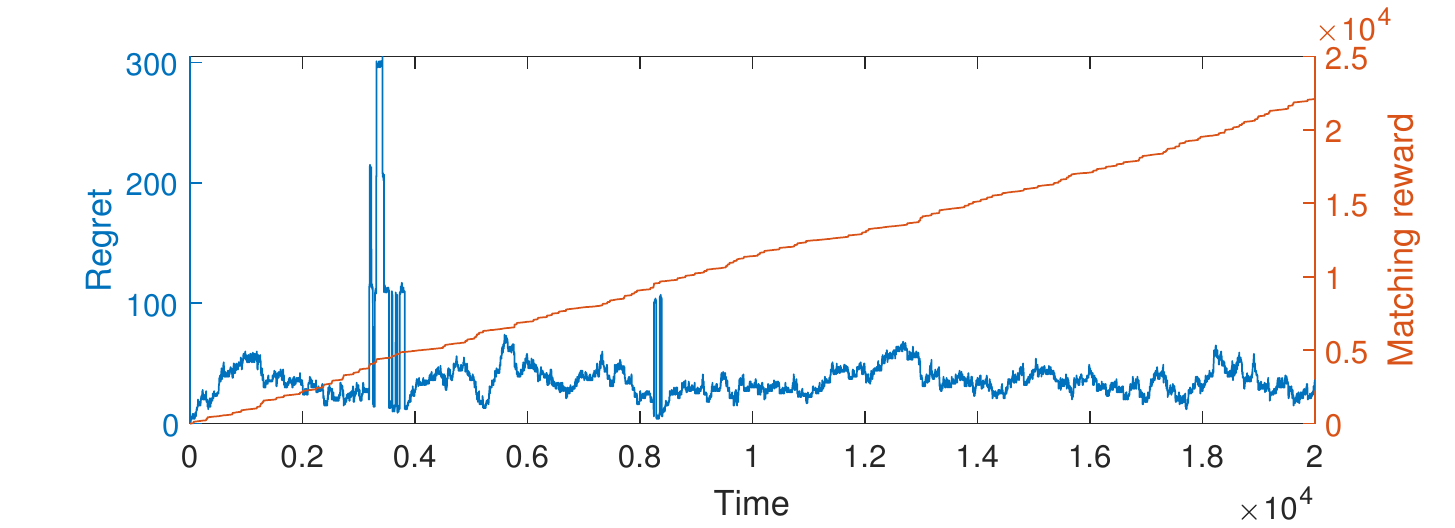}}
\subfigure{\includegraphics[width=3.2in]{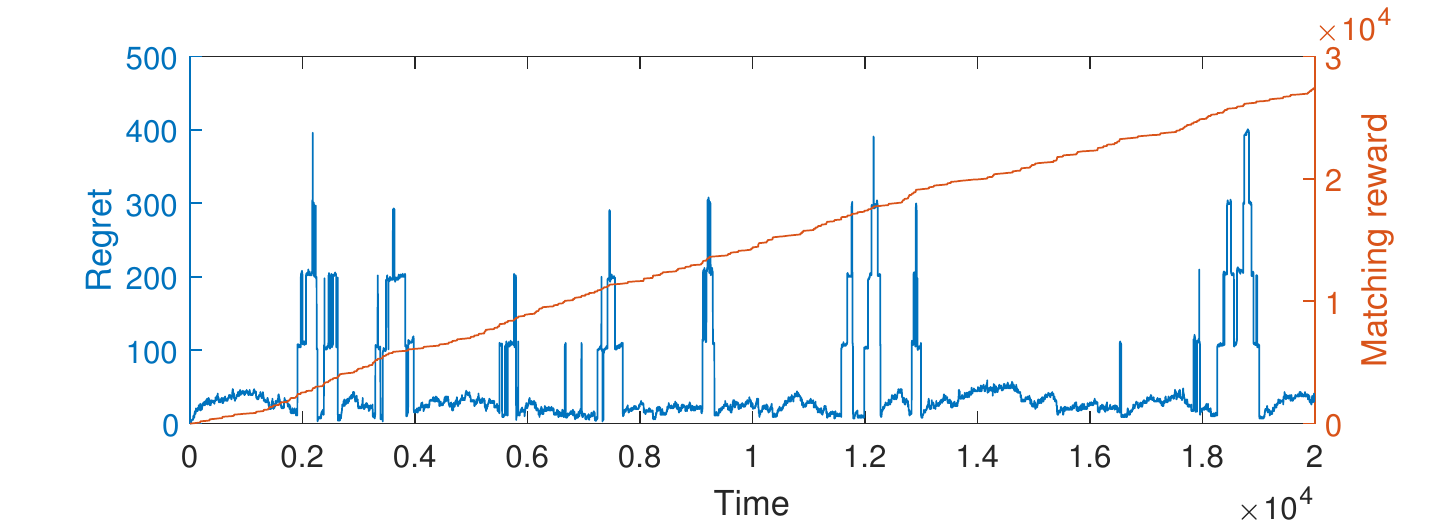}}
\hspace{-0.1in}
\subfigure{\includegraphics[width=3.2in]{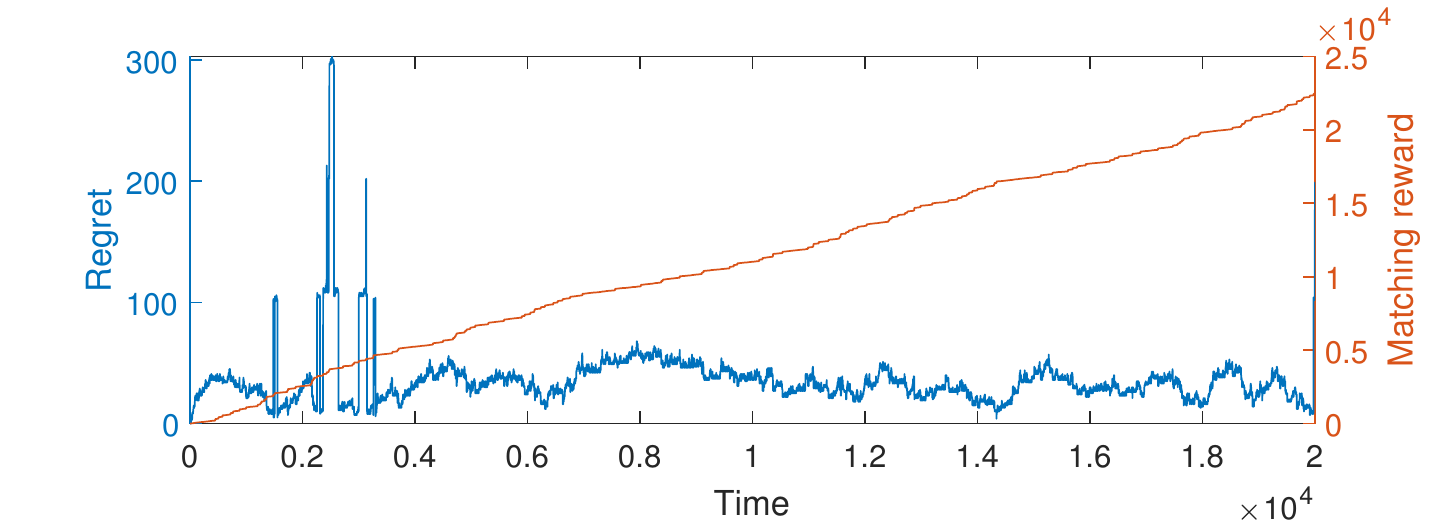}}
\caption{Four simulation paths for the Sum-of-Squares algorithm applied to the instance in Figure~9 of \citep{kerimov2021dynamic} (reproduced in Figure~\ref{fig:multiway_instance}). The red curves show the cumulative matching reward for the offline optimal policy. The blue curves show the regret of the proposed Sum-of-Squares algorithm.}
\label{fig:Example_Fig9}
\end{center}
\end{figure}

{\bf Instance 1:} The first instance in our experiments is an instance with online-queueable resources which we borrow from  \cite{kerimov2021dynamic} and reproduce in Figure~\ref{fig:multiway_instance}. The instance was mentioned in that work as one where a natural greedy policy does not give bounded regret. The intuition behind the observation is that the match of type 1 generates the highest rewards, but involves resources which are part of other matching configurations. A greedy algorithm may not leave enough resources for type 1 matches. This is where the idea that the right view of designing a greedy algorithm is that it should be greedy in committing resources to configurations, rather than only at the time of matching is critical. A second notable feature of the experiments in \cite{kerimov2021dynamic} is that depending on the choice of the tuning parameter which controls the frequency of the resolving algorithm in \cite{kerimov2021dynamic}, the optimality gap could be potentially quite large. Figure~\ref{fig:Example_Fig9} shows one simulation run of the Sum-of-Squares algorithm for this instance from \cite{kerimov2021dynamic} which highlights the bounded regret of the proposed algorithm.

\begin{figure}[ht!]
\begin{center}
\subfigure{\includegraphics[width=3.2in]{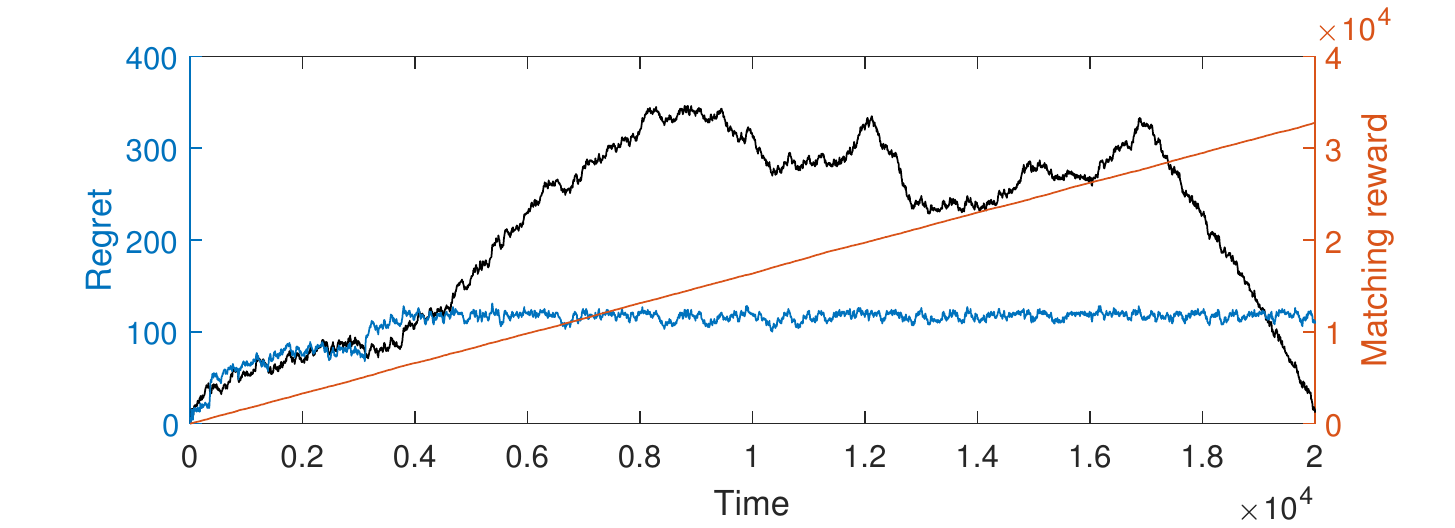}}
\hspace{-0.1in}
\subfigure{\includegraphics[width=3.2in]{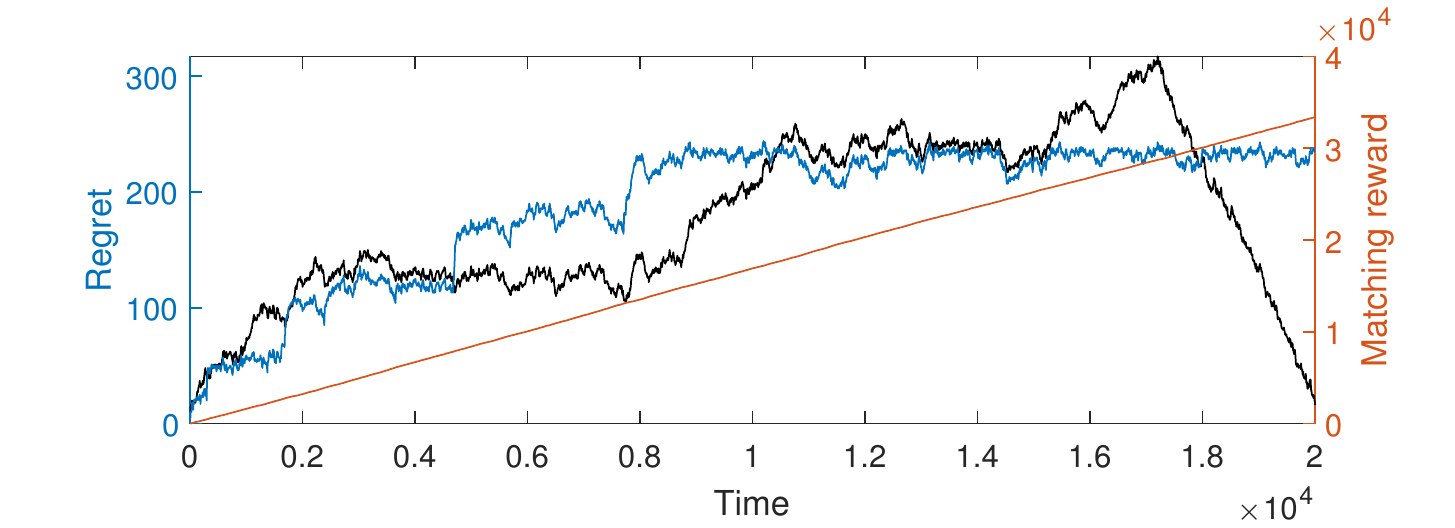}}
\subfigure{\includegraphics[width=3.2in]{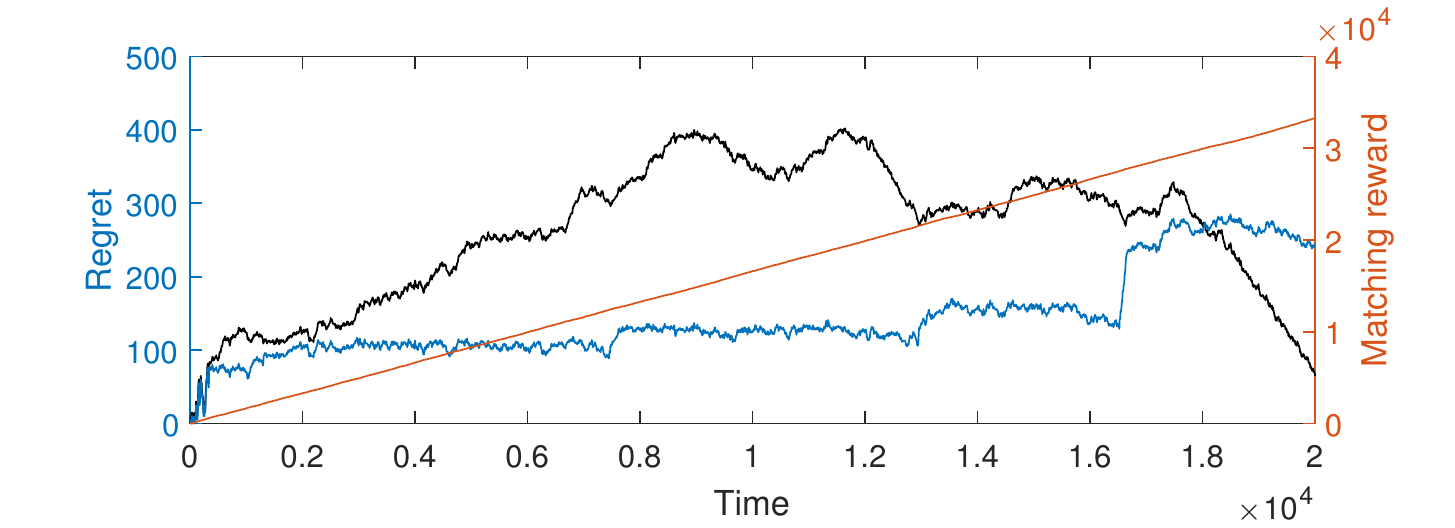}}
\hspace{-0.1in}
\subfigure{\includegraphics[width=3.2in]{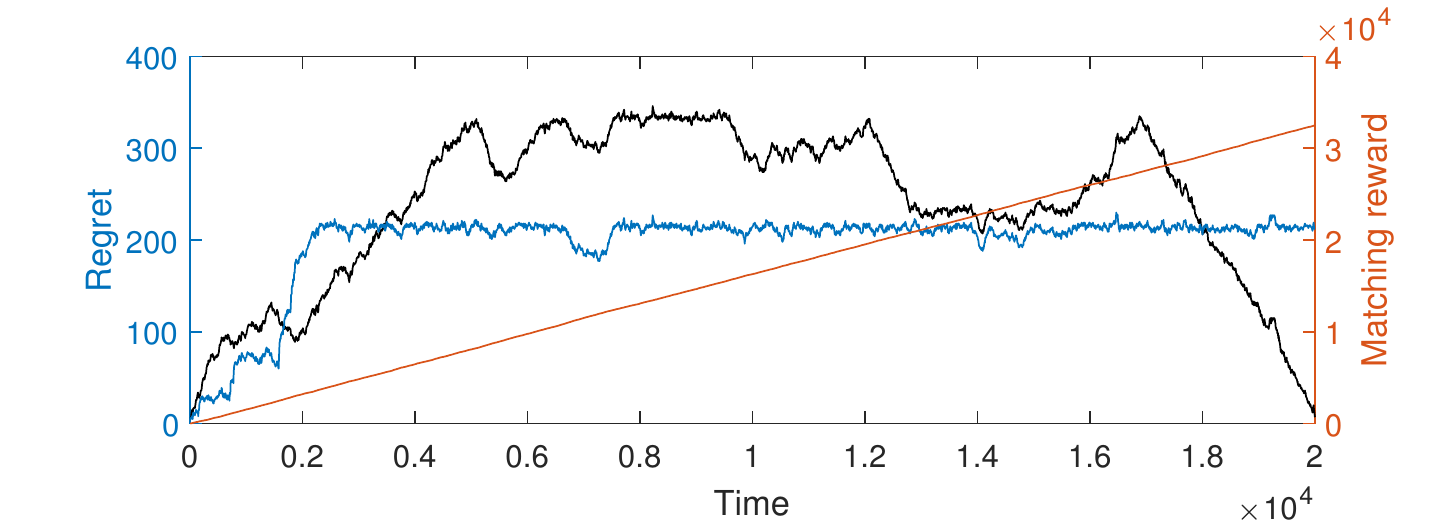}}
\caption{Four simulation paths for the Sum-of-Squares algorithm applied to the instance in Figure~\ref{fig:LB_instance}. The offline optimal computed here is an upper bound on the matching reward under the assumption that all arrivals can be queued and rematched. The black curve shows the regret of the Dynamic Programming based policy for maximizing expected reward at the terminal time. The blue curve shows the regret of the Sum-of-Squares algorithm.}
\label{fig:Example_LB}
\end{center}
\end{figure}

{\bf Instance 2:} The second instance we simulate is the lower bound instance from Figure~\ref{fig:LB_instance} used to prove Theorem~\ref{thm:mainLB} for the case where configurations have both online-queueable and online-nonqueueable resources. In Figure~\ref{fig:Example_LB} we compare our algorithm (blue curve) against the offline optimal (which is allowed to rematch all arrivals, even if the matches may not be realizable online), and the performance of a Dynamic Programming optimal policy (black curve) which aims to maximize the terminal matching reward. While the terminal regret of the DP optimal policy is superior to the terminal regret of the Sum-of-Squares algorithm, Sum-of-Squares performs comparably to the DP optimal policy for the entire interval $[T]$ as Theorems~\ref{thm:mainUB} and \ref{thm:mainLB} prove.

{\blue {\bf Instance 3:} The third instance we simulate is for the stochastic bin packing problem, where our goal is to compare the performance of our $\SoS$ algorithm compared to the Sum-of-Squares algorithm of \cite{csirik2006sum}, and the modification where dead-end levels are forbidden. To be able to use the algorithm of \cite{csirik2006sum} as is, we will focus on a packing instance corresponding to Bounded Waste distribution (which are a subclass of distributions exhibiting $\GPG$). In the instance we simulate (adapted from \citep{csirik2006sum}), items of size $2$ or $3$ arrive over the packing horizon, and must be packed on arrival into bins of size $B=9$. The size of the arriving item is sampled i.i.d., with $\lambda_2$ denoting the probability that the item is of size 2, and $\lambda_3=1-\lambda_2$ that it is of size 3. Any distribution with $\lambda_3 \geq 1/4$ is Perfectly Packable, and is a Bounded Waste distribution if the inequality is strict.

\begin{figure}[ht!]
\begin{center}
\subfigure{\includegraphics[width=6.5in]{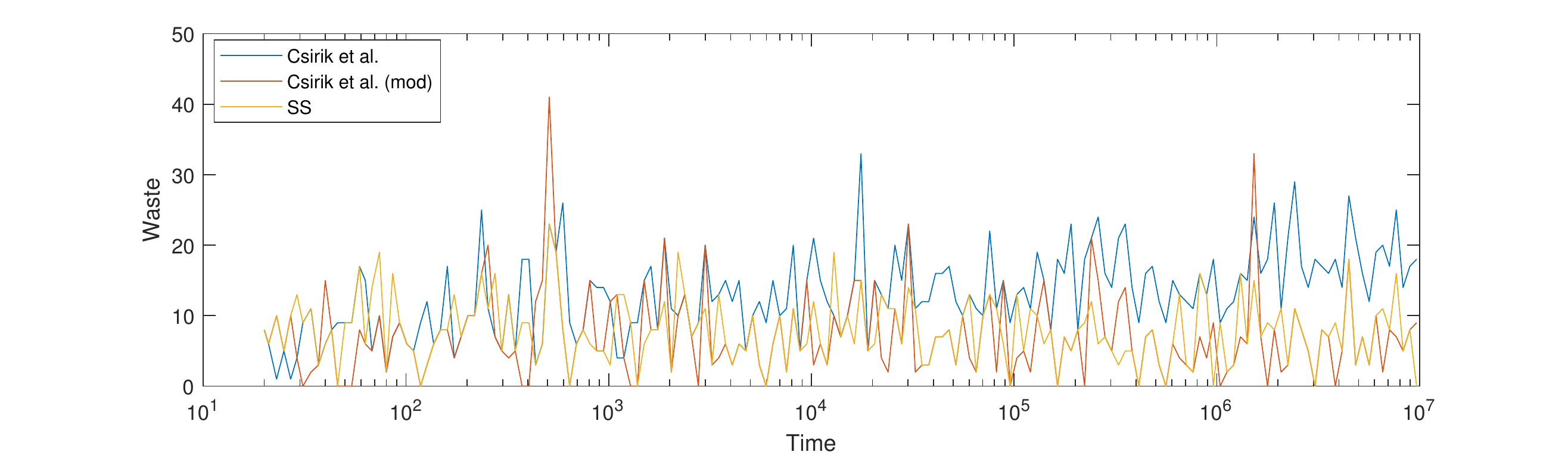}}
\subfigure{\includegraphics[width=6.5in]{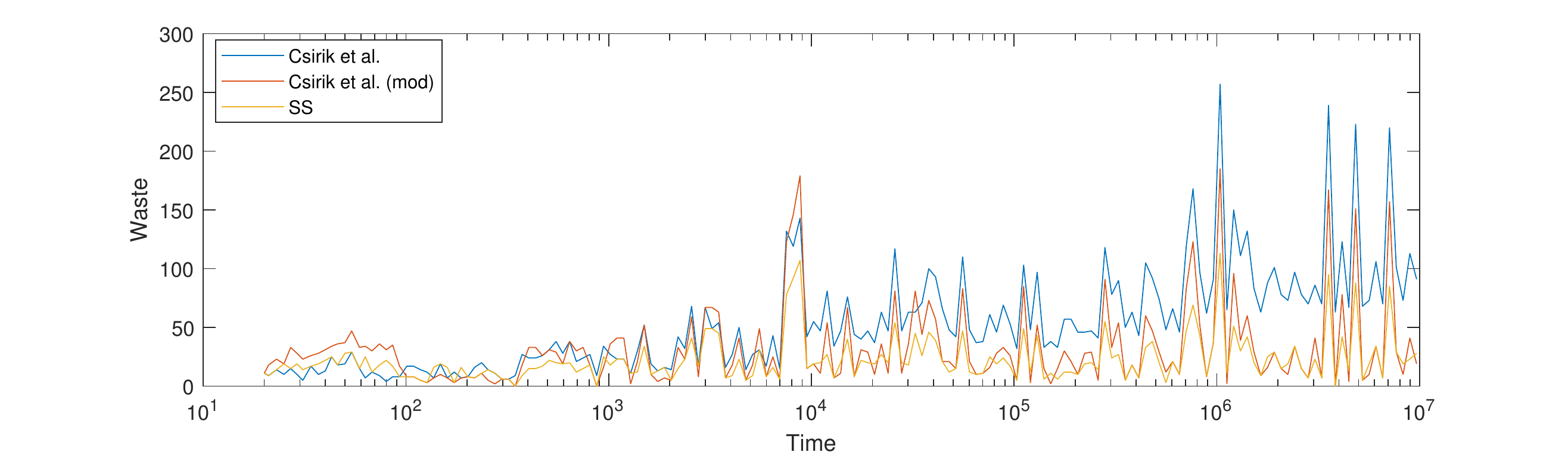}}
\caption{Simulation paths for the Sum-of-Squares algorithm applied to  a stochastic bin packing instance with Bounded Waste item size distribution. The bin size is $B=9$, and the item size distribution is $1-\lambda_2 = \lambda_3 = 1/4 + \epsilon$, with $\epsilon = 1/8$ for the top plot and $\epsilon = 1/64$ for the bottom plot. The curves show the wasted space in opened bins used by three algorithms: \textit{SS} corresponds to the algorithm proposed in the current paper, \textit{Csirik et al.} corresponds to the basic Sum-of-Squares algorithm of \cite{csirik2006sum}, and \textit{Csirik et al. (mod)} corresponds to the variation where bins of level 8 (a dead-end level) are never created. }
\label{fig:Example_bin_packing}
\end{center}
\end{figure}

Figure~\ref{fig:Example_bin_packing} shows the simulation results, where we plot the sample path of the waste incurred by the algorithm of \cite{csirik2006sum} and our proposed $\SoS$ algorithm. Waste stands for the unused space in the open bins used by the online packing algorithms. The solution to the $\SPP$ would correspond to a waste of 0, so our plots are an upper bound on the regret. We compare our algorithm with the basic Sum-of-Squares algorithm of \cite{csirik2006sum} (labeled \textit{Csirik et al.} in the plots), and the modification where dead-end levels are never created (labeled \textit{Csirik et al. (mod)}). For the instance we simulate, level 8 is the only dead-end level. We simulate two distributions: with $\lambda_3 = 1/4 + 1/8$ (top) and $\lambda_3 = 1/4 + 1/64$. As we mentioned earlier, we get a Bounded Waste distribution when $\lambda_3 > 1/4$, and the distance to $1/4$ is a measure of the general position gap. 

The first observation we can make is that the basic algorithm of \cite{csirik2006sum} incurs a waste that grows as $\Theta(\log t)$, as proved by the authors. This is more apparent in the bottom plot, and occurs due to accumulation of level 8 bins. The second observation is that both, our $\SoS$ algorithm, as well as the modified Sum-of-Squares algorithm of \cite{csirik2006sum}, have waste which does not seem to grow with $t$, and have comparable performance.}
\section{Concluding Remarks}

The main purpose of this research note is to highlight the power of knowing partial information, i.e., the basic feasible activities, about the optimal offline solution. While some of the ideas in this paper have perhaps been hidden in some of the existing literature, we believe they have not been expressed explicitly. A second contribution we make is to present a slightly more unified view of online multiway matching by considering three types of resources. Finally, the potential function analysis we conduct for non-stationary distributions by borrowing some tricks from amortized analysis is likely to be of broader interest. We would like to reiterate that for some of the problems which fall within our unified framework, stronger results are known (bounded regret without GPG), and whether these results can be obtained via pure greedy algorithms would be an interesting direction to explore. Further, whether knowledge of basic feasible activities alone can be used to create priority policies which are oblivious to the entries of matrix $\Mb$ (other than using whether they are zero or non-zero, as is done by \cite{stolyar2013tightness} for queueing systems) is interesting to explore.


\theendnotes

\bibliographystyle{ormsv080} 
\bibliography{references} 

%
\newpage

\begin{APPENDICES}
\section{Proofs}
\subsection{Proof of Lemma~\ref{lem:icx_bound}}

First observe that for any random variable $\Delta$ with support on $[-K_1, +K_1]$ and mean $\expct{\Delta}$, the random variable $\Delta'$ with support on $\{-K_1, +K_1\}$ and mean $\expct{\Delta'} \geq \expct{\Delta}$ satisfies $\Delta \leqicx \Delta'$.

To prove the lemma, we give a construction that satisfies the second part of Definition~\ref{defn:icx}. That is, we construct $\{\hat{\Upsilon}^t\}$ and $\{\hat{\Gamma}^t\}$ on a common probability space so that $\expct{\hat{\Gammab} | \hat{\Upsilonb}} \geq \hat{\Upsilonb}$ almost surely. In our construction $\hat{\Upsilon}^t = \Upsilon^t$ for all $t \in [T]$, and the only interesting part is the construction of $\{\hat{\Gamma}^t\}$ on the probability space of $\{\Upsilon^t\}$.

{\bf Base case:} $\hat{\Gamma}^\tau = \Gamma^\tau$. By the choice of $\Gamma^\tau$, we have $\hat{\Gamma}^\tau \geq \hat{\Upsilon}^\tau$ almost surely.

{\bf Inductive Step:} Assume that we have constructed on a joint probability space $\{\hat{\Upsilon}^s\}_{s\in[\tau,t]}$ and $\{\hat{\Gamma}^s\}_{s\in[\tau,t]}$ satisfying $\expct{\hat{\Gammab} | \hat{\Upsilonb}} \geq \hat{\Upsilonb}$, obeying the marginals. Denote
\[ \hat{\Delta}_{\Upsilon}^{t+1}  = \hat{\Upsilon}^{t+1} - \hat{\Upsilon}^{t}.\]
The random variable $\hat{\Gamma}^{t+1}$ is given by:
\begin{align*}
    \hat{\Gamma}^{t+1} &= 
    \begin{cases}
        \max\left\{ \Gamma_{min} , \hat{\Gamma}^{t} + \hat{\Delta}_\Gamma^{t+1}  \right\} & \mbox{if } \hat{\Upsilon}^{t} \geq K_2, \\
        \max\left\{ \Gamma_{min} , \hat{\Gamma}^{t} + {\Delta}_\Gamma^{t+1}  \right\} & \mbox{if } \hat{\Upsilon}^{t} < K_2.        
    \end{cases}
\end{align*}
In the above  $\hat{\Delta}^{t+1}_\Gamma$ is a random variable with support $\{-K_1, +K_1\}$, mean $\expct{\hat{\Delta}^{t+1}_\Gamma} = -2\delta' K_1$, that is coupled with $\hat{\Delta}^{t+1}_{\Upsilon}$ so that \textit{(i)} conditioned on $\hat{\Delta}^{t+1}_{\Upsilon}$, $\hat{\Delta}^{t+1}_\Gamma$ is independent of $\{\hat{\Gamma}^s\}_{s\in [t]}$ and $\{\hat{\Upsilon}^s\}_{s\in [t]}$, and \textit{(ii)} and
\[ \expct{\hat{\Delta}^{t+1}_{\Gamma} | \hat{\Delta}^{t+1}_{\Upsilon}} \geq \hat{\Delta}^{t+1}_{\Upsilon}. \]
The random variable $\Delta_{\Gamma}^{t+1} = \xi^{t+1} \gamma$ with $\xi^{t+1} \in \{-1,+1\}$ denoting a random variable with distribution given in the Lemma, and independent of $\hat{\Upsilonb}$ and $\{\hat{\Gamma}^s\}_{s\in [t]}$. 

Note that the marginal distributions of $\Delta_\Gamma^{t+1}$ and $\hat{\Delta}_\Gamma^{t+1}$ are identical, and both random variables are independent of $\{\hat{\Gamma}_{s}\}_{s\in [t]}$, thus proving that the marginal distribution of $\hat{\Gammab}$ agrees with $\Gammab$.

It remains to show that we have
\[ \expct{\hat{\Gamma}^{t+1} | \hat{\Upsilonb}} = \expct{\hat{\Gamma}^{t+1} | \{\hat{\Upsilon}^s\}_{s\in[t+1]}} \geq \hat{\Upsilon}^{t+1}. \]

On the event $\{\hat{\Upsilon}^{t} < K_2\}$ we have $\hat{\Upsilon}^{t+1} \leq \hat{\Upsilon}^t + K_1 \leq K_2 + K_1 \leq \Gamma_{\min} \leq \hat{\Gamma}^{t+1}$.

On the event $\{\hat{\Upsilon}^{t} \geq K_2\}$:
\begin{align*}
    \expct{ \hat{\Gamma}^{t+1} | \{\hat{\Upsilon}^s\}_{s\in [t+1]} , \hat{\Upsilon}^t \geq K_2} & \geq      \expct{ \hat{\Gamma}^{t} | \{\hat{\Upsilon}^s\}_{s\in [t+1]} , \hat{\Upsilon}^t \geq K_2 } +     \expct{ \hat{\Delta}_\Gamma^{t+1} | \{\hat{\Upsilon}^s\}_{s\in [t+1]} , \hat{\Upsilon}^t \geq K_2 } \\
    & \geq \hat{\Upsilon}^{t} +  \expct{ \hat{\Delta}_\Gamma^{t+1} | \hat{\Delta}^{t+1}_\Upsilon , \hat{\Upsilon}^t \geq K_2 } \\
    & \geq \hat{\Upsilon}^{t} + \hat{\Delta}^{t+1}_\Upsilon \\
    &= \hat{\Upsilon}^{t+1},
\end{align*}
as required. The first inequality follows since $\hat{\Gamma}^{t+1} = \max\{ \Gamma_{\min}, \hat{\Gamma}^t + \hat{\Delta}^{t+1}_\Gamma\}$, the second follows by the induction hypothesis $ \expct{\hat{\Gamma}^{t} | \{\hat{\Upsilon}^s\}_{s\in[t+1]}} \geq \hat{\Upsilon}^{t}$, and the third inequality follows by the coupling between $\hat{\Delta}^{t+1}_\Upsilon$ and $\hat{\Delta}^{t+1}_\Gamma$.

\subsection{Proof of Lemma~\ref{lem:bounded_discrepancy}}
It is an easy observation that for each $t\in [T]$, $\sum_{i \in \Ical^\off \cup \{0\}} \left( A^t_i - \bar{\lambda}_i \cdot t \right) = \sum_{i \in \Ical^\off \cup \{0\}} A^t_i - t = 0 $. Therefore, if for some $i$, $A^{t-1}_i - \bar{\lambda}_i \cdot (t-1) > 0$, then there exists another $i'$ with $A^{t-1}_{i'} - \bar{\lambda}_i \cdot (t-1) < 0$, and therefore
\[ A^{t}_i - \bar{\lambda}_i \cdot t < A^{t-1}_i - \bar{\lambda}_i \cdot (t-1) .\]
Since $A^{t}_i - \bar{\lambda}_i \cdot t$ increases by at most $1 - \bar{\lambda}_i$ in any time step, and decreases whenever it is strictly positive, it takes a maximum value of $1 - \bar{\lambda}_i$.

To lower bound $A_i^t$, we note that $A_i^t = t - \sum_{i' \in \Ical^\off \cup \{0\} \setminus \{i\} } A_{i'}^t \geq t - \sum_{i' \in \Ical^\off \cup \{0\} \setminus \{i\} } ( \bar{\lambda}_{i'} \cdot t + 1  - \bar{\lambda}_{i'}) \geq \bar{\lambda}_{i} \cdot t - n^\off $.  

Finally, $A_i^T = \bar{\lambda}_i \cdot T$ follows because $A_i^T$ is at most $ \bar{\lambda}_i \cdot T + (1-\bar{\lambda}_i)$, and hence being integral, at most $\bar{\lambda}_i \cdot T$. Combining $\sum_i A_i^T = T$, $A_i^T \leq \bar{\lambda}_i T$ for all $i \in \Ical^\off \cup \{0\}$, and $\sum_i \bar{\lambda}_i T = T$, we get $A_i^T = \bar{\lambda} \cdot T$.  

\subsection{Proof of Proposition~\ref{prop:GPG_dual}}
We first show that Definition~\ref{def:GPG} implies uniqueness of the optimal solution $\alphab^*$ of \eqref{mp:DSPP}. Let $\Mcal_+$ be an optimal basis. By the assumption in Definition~\ref{def:GPG}, the optimal primal variables for $\hat{\lambdab} \in \Bcal_{\epsilon, \TV}(\lambdab)$ with the basis $\Mcal_+$ are given by:
\[ \xb^*_{\Mcal_+}(\hat{\lambdab}) = \Mb_{\Mcal_+}^{-1} \hat{\lambdab},\]
with optimal objective function value
\[ \rb_{\Mcal_+}^\top \Mb_{\Mcal_+}^{-1} \hat{\lambdab} .\]
 It follows that $\alphab^* = \left(\rb_{\Mcal_+}^\top \Mb_{\Mcal_+}^{-1}\right)^\top$ is an optimal solution to the dual $\DSPP(\hat{\lambdab})$ (see, e.g., the lecture notes by \cite{lavrov2020}). Since $\alphab^*$ is an optimal dual solution for all $\hat{\lambdab} \in \Bcal_{\epsilon,\TV}(\lambdab)$, and observing  that we wrote the dual program \eqref{mp:DSPP} as
 \[ \max_{\alphab \in \Acal} \  \alphab^\top \lambdab \]
for a polyhedron $\Acal$ which depends on $\rb$ and $\Mb$ but not on $\lambdab$, $\alphab^*$ must be a vertex of $\Acal$, and hence the unique optimal solution to \eqref{mp:DSPP}.\\
Conversely, since $\alphab^*$ is an optimal solution for all $\hat{\lambdab} \in \Bcal_{\epsilon, \TV}(\lambdab)$, pick any $|\Ical|$ linear independent constraints that bind at $\alphab^*$, and call these the basis $\Mcal_+$. It follows that $\xb^*_{\Mcal_+}(\hat{\lambdab}) = \Mb_{\Mcal_+}^{-1} \hat{\lambdab}$ is an optimal solution to $\DSPP(\hat{\lambdab})$.

\subsection{Proof of Lemma~\ref{lem:gap_virtual_true}}

{\bf Case 1:} For a configuration $m$ with online-nonqueueable resources, a virtual match happens whenever an online-nonqueueable resource is committed to the configuration. However, a true match only happens if the offline resources are available (globally), and online-queueable resources are available (locally, per the true queue length). 

To account for the gap in matches due to lack of online-queueable resources, we will use the difference between the true and virtual queue lengths for such resources. More formally, let $i \in \Ical(m) \cap \Ical^\onnq$. Then the virtual queue length $\Qs_{im}$ decreases by $1$ whenever a virtual match happens, and the true queue length $Q_{im}$ decreases by $1$ only when a true match happens. However, both $\Qs_{im}$ and $Q_{im}$ increase when a unit of resource $i$ is committed to $m$. Therefore,
\[ \Xs^t_m - X^t_m = Q^t_{im} - \Qs^{t}_{im}. \]
Now consider the last time before $t$ that a virtual match happened but a true match did not happen because of lack of an online-queueable resource. Denote this time by $\theta$, and let us continue to denote by $i$ the type of this online-queueable resource. Therefore we must have $Q^{\theta}_{im} = Q^{\theta-1}_{im} \leq 0$, and $\Qs^{\theta}_{im} = \Qs^{\theta-1}_{im} - 1 $. Therefore, the gap between virtual and online matches up to and inclusive of time $\theta$ is
\[ \Xs^\theta_m - X^\theta_m = Q^{\theta}_{im} - \Qs^{\theta}_{im} \leq 1 - \Qs^{\theta}_{im}  \leq 1 + \max_{s \in [t]}  \left( \Qs^{s}_{im} \right)^- \leq 1 + \max_{i \in \Ical(m) \cap \Ical^\onnq} \max_{s \in [t]}  \left( \Qs^{s}_{im} \right)^-. \]
The gap in matches due to time interval $[\theta+1,t]$ is solely due to lack of offline resources globally, which we bound next. 
We first observe that the total number of units of a resource $i \in \Ical^\off$ used by all configurations can be bounded by the units consumed due to virtual matches
\[ \sum_{m \in \Mcal_+} M_{im} X^t_m  \leq \sum_{m \in \Mcal_+} M_{im} \Xs^t_m.   \]
Next, let us view $Y_{im}^t$ as the ``quota'' of resource $i$ that is assigned to configuration $m$. If all configurations consume less units of resource $i$ than their quota, then there would be no lost true matches caused by lack of resource type $i$, and in this case all the virtual queues $\Qs_{im}^t$ would be non-negative. Therefore, we can upper bound the gap between virtual and true matches of configuration $m$ due to resource $i$ into \textit{(i)} the excess virtual consumption of resource $i$ by $m$ beyond its own quota -- this overconsumption is bounded from above by $\left(\Qs_{im}^t\right)^-$; and \textit{(ii)} the over-consumption of resource $i$ by other configurations, thereby eating into configuration $m$'s quota -- this over-consumption is upper bounded by
\[ \sum_{ m': i \in \Ical(m), m \neq m' }  \left( \Qs_{im'}^t \right)^-. \]
This in turn leads to an upper bound on the gap between true and virtual matches of configuration $m$ due to resource $i \in \Ical^\off$ of 
\[ 1 + \sum_{ m': i \in \Ical(m)}  \left( \Qs_{im'}^t \right)^-. \]
Summing the gaps due to online-queueable and offline resources, we get the bound in the Lemma:
\[ \Xs_m^t - X_m^t \leq  \max_{ i \in \Ical(m) \cap \Ical^\onq} \max_{s \in [t] } \left(\Qs_{im}^s\right)^- + \sum_{ i \in \Ical(m) \cap \Ical^\off } \sum_{ m' | i \in \Ical(m')} \left(\Qs_{im'}^t\right)^- + |\Ical(m)|.  \]

{\bf Case 2:} If configuration $m$ does not have online-nonqueueable resources, then the gap between virtual and true matches occurs only due to lack of offline-resources (since virtual match is not performed if there are not enough virtual online-queueable resources, and the true queue length is always larger than virtual queue length for these resources). The gap due to lack of offline resources can be bound as in Case 1.

If the configuration $m$ does not have online-queueable resources, then again a virtual match is performed as soon as the online-nonqueueable resource arrives, and the gap between virtual and true matches is only due to lack of offline resources. This again can be bounded as in Case 1.

\subsection{Proof of Lemma~\ref{lem:gap_opt_virtual}}

The final virtual queue lengths for all the resources are given by $\sum_m \Qsb^t_m$, and therefore the quantity of resources consumed to generate the virtual matches are $\Ab^t - \sum_m \Qsb^t_m$. Since we only generate matches in the set $\Mcal_+$, and by our general position gap assumption $\Mb_{\Mcal_+}$ is full rank, the number of virtual matches is given by the unique solution to the following linear system:
\[    \forall i \in \Ical \ : \ \sum_{m \in \Mcal} M_{im} \cdot \Xs^t_m = A_i^t - \sum_{m} \Qs_{im}^t, \]
or,
\[ \Xsb^t = (\Mb_{\Mcal_+})^{-1} \left( \Ab^t - \sum_{m} \Qsb_{m}^t \right), \]
or in terms of matching reward,
\begin{align}
\label{eqn:regret_decomp}
\rb^\top_{\Mcal_+}\Xsb^t = \rb^\top_{\Mcal_+} (\Mb_{\Mcal_+})^{-1} \left( \Ab^t - \sum_{m} \Qsb_{m}^t \right).  
\end{align}
Assuming $\Ab^t/t \in \Bcal_{\epsilon_0,\TV}(\bar{\lambdab})$, $\GPG_{\epsilon_0}$ condition implies, $\rb^\top_{\Mcal_+} (\Mb_{\Mcal_+})^{-1} \Ab^t = \OPT(\Ab^t)$, and therefore,
\[ \sum_m r_m \cdot \Xs^t_m = \OPT(\Ab^t) - \rb^\top_{\Mcal_+} (\Mb_{\Mcal_+})^{-1}  \sum_{m} \Qsb_{m}^t. \]

\subsection{Proof of Lemma~\ref{lem:bounded_regret}}
Starting from \eqref{eqn:regret_decomp},
\begin{align}
\label{eqn:regret_decomp2}
    \expct{\sum_{m \in \Mcal_+} r_m \Xs_m^t} &= \rb^\top_{\Mcal_+} (\Mb_{\Mcal_+})^{-1} \expct{\Ab^t} - \rb^\top_{\Mcal_+} (\Mb_{\Mcal_+})^{-1} \expct{\sum_{m} \Qsb_{m}^t}.
\end{align}
Our goal is to relate $ \rb^\top_{\Mcal_+} (\Mb_{\Mcal_+})^{-1} \expct{\Ab^t}$ and $\expct{\OPT(\Ab^t)}$. In the rest of the proof we will consider $t \geq \tau+1$, otherwise the Lemma is straightforward since all the terms involved are $\Ocal(\tau)$.  
Using the dual $\eqref{mp:DSPP}$ of $\eqref{mp:SPP}$
it is easy to see that $\OPT(\lambdab)$ is a concave, piece-wise linear, and Lipschitz function. Therefore,
\begin{align}
    \nonumber
    \expct{\OPT(\Ab^t) } & \leq \OPT(\expct{\Ab^t}) \\
    \nonumber
    & = \OPT\left( \sum_{s=1}^t \lambdab^s \right) \\
    \nonumber
    &= \OPT\left( \sum_{s=1}^{t-\tau} \hat{\lambdab}^s \right) + \left[ \OPT\left( \sum_{s=1}^t \lambdab^s \right)  - \OPT\left( \sum_{s=1}^{t-\tau} \hat{\lambdab}^s \right) \right] \\
    \nonumber
    &= \rb^\top_{\Mcal_+} (\Mb_{\Mcal_+})^{-1} \sum_{s=1}^{t-\tau} \hat{\lambdab}^s +  \left[ \OPT\left( \sum_{s=1}^t \lambdab^s \right)  - \OPT\left( \sum_{s=1}^{t-\tau} \hat{\lambdab}^s \right) \right] \\
    \nonumber
    &= \rb^\top_{\Mcal_+} (\Mb_{\Mcal_+})^{-1} \sum_{s=1}^{t} {\lambdab}^s + \rb^\top_{\Mcal_+} (\Mb_{\Mcal_+})^{-1} \left[
      \sum_{s=1}^{t-\tau} \hat{\lambdab}^s - 
    \sum_{s=1}^{t}{\lambdab}^s\right] +  \left[ \OPT\left( \sum_{s=1}^t \lambdab^s \right)  - \OPT\left( \sum_{s=1}^{t-\tau} \hat{\lambdab}^s \right) \right]\\
    &= \rb^\top_{\Mcal_+} (\Mb_{\Mcal_+})^{-1}\expct{\Ab^t} + \Ocal(\tau)
    \label{eqn:regret_decomp3}
\end{align} 
where in the third step we have used the smoothed $\GPG$ assumption on $\{\lambdab^t\}$, in the fourth step we have used $ \frac{1}{t-\tau} \sum_{s=1}^{t-\tau} \hat{\lambdab}^s \in \Bcal_{\epsilon_0,\TV}(\bar{\lambdab})$, and in last step we have used $\sum_{s=1}^{t-2\tau} \hat{\lambdab}^s \leq \sum_{s=\tau+1}^{t} {\lambdab}^s  \leq  \sum_{s=1}^{t} \hat{\lambdab}^s $, and that $\OPT$ is Lipschitz. Combining \eqref{eqn:regret_decomp2} and \eqref{eqn:regret_decomp3},
\[ 
    \expct{\sum_{m \in \Mcal_+} r_m \Xs_m^t} = \expct{\OPT(\Ab^t)} - \rb^\top_{\Mcal_+} (\Mb_{\Mcal_+})^{-1} \expct{\sum_{m} \Qsb_{m}^t} - \Ocal(\tau).
    \]

\subsection{Proof of Lemma~\ref{lem:drift_lemma}}

Our proof combines the Lyanunov drift analysis from Section~\ref{sec:iid} and amortized analysis to handle non-stationary arrival distributions. We begin with a recap of the proof from Section~\ref{sec:iid}, where this analysis fails, and how the amortized analysis idea helps.

We will use the following notation throughout: for a stochastic process $f^t$ (think of this as a Lyapunov/potential function), let 
\[ \expctsub{\lambdab , \pi}{\Delta^t f} := \expctsub{I^{t} \sim \lambdab , \pi}{f^{t} - f^{t-1} | \Fcal^{t-1}} \]
be a shorthand for the expected change in $f$ from time $t-1$ to $t$ if the type of the resource arriving at time $t$ is sampled from distribution $\lambdab$ and ``placed'' according to policy $\pi$. By placing according to $\pi$, we mean the choice of $m^t$ in Algorithm~\ref{alg:SoS} is replaced by policy $\pi$. The policy $\pi$ can be randomized, as well as can depend on the entire history until time $t$ of the arrivals and its own randomized actions. We will us $\SoS$ to denote the policy in Algorithm~\ref{alg:SoS}.

A particular randomized policy that will be key in the analysis is the policy $\Rcal(\lambdab)$ that we mentioned in the proof of Lemma~\ref{lem:drift_main_illustration}.
\begin{definition}[Randomized policy $\Rcal(\lambdab)$]
\label{def:Rlambda} 
Suppose the item arriving at time $t$ is sampled from a distribution $\lambdab \in \Bcal_{\epsilon_0,\TV}(\bar{\lambdab})$. The randomized policy, $\Rcal(\lambdab)$, first solves $\SPP(\lambdab)$ with a given optimal basis $\Mcal_+$. Let $\xb$ denote the unique optimal solution with basis $\Mcal_+$. Given the type of the arrival $I^t=i$, policy $\Rcal(\lambdab)$ commits the item to configuration $m$ with probability $\frac{M_{im}x_m}{\sum_{m'} M_{im'}x_{m'}}$. That is, the resource is committed to a configuration with probability distribution consistent with the solution $\xb$ to $\SPP(\lambdab)$.
\end{definition}

\subsubsection{Proof Sketch}
\ \\ 
Let us first attempt to reproduce the analysis in Section~\ref{sec:iid} for the general case (non-stationary arrivals, all three types of resources). Consider the Lyapunov function $f^t = \sqrt{\Phi(\Qsb^t)}$. We first proved that since the queue lengths can change by $\pm 1$ in any one time step, the change in this potential function is bounded by $\sqrt{2}$. This step still holds for the general case, since the virtual queues also change by $\pm 1$.  

Next, to prove expected decrease condition, we meed to bound $\expctsub{\lambdab^t, \SoS}{\Delta^t f}$ when $f^{t-1}$ is large. This is done by constructing a probabilistic policy, $\pi^t$, which is easier to study and upper bounds the drift for policy $\SoS$ since for any policy $\pi^t$, $\expctsub{\lambdab^t, \SoS}{\Delta^t f} \leq \expctsub{\lambdab^t, \pi^t}{\Delta^t f}$.

To construct $\pi^t$, we first need to prove that if $\Phi(\Qsb^{t-1})$ for the current state is large, then there must exist some resource type $i^*$, such that if this item is packed by $\SoS$ then the change in the sum-of-squares potential is proportional to $-\sqrt{\Phi(\Qsb^{t-1})}$ (Lemma~\ref{lem:negative_drift}). Next, one show that for any $\mathring{\lambdab} \in \Bcal_{\epsilon_0,\TV}(\bar{\lambdab})$, the randomized policy $\Rcal(\mathring{\lambdab})$ gives an expected change in the Sum-of-Squares potential is bounded above by a constant (Lemma~\ref{lem:expected_inc}). 
 The policy $\pi^t$ is then constructed by splitting 
\[ \lambdab^t = \delta \eb_{i^*} + (1-\delta) \underbrace{\left( \frac{\lambdab^t - \delta \eb_{i^*}}{1-\delta} \right)}_{=: \mathring{\lambdab}^t},\]
for some $\delta > 0$. 
We now bound the drift under $\SoS$ by arguing as follows: with probability $(1-\delta)$ the item distribution is $\mathring{\lambdab}^t$ which can be packed using $\Rcal(\mathring{\lambdab}^t)$ so that the expected change in the sum-of-squares potential is $\Ocal(1)$, and with probability $\delta$ the item distribution is $\eb_i$ which can be packed greedily with a large negative change in the potential. We call this hybrid randomized+greedy policy $\Hyb(\lambdab^t,\Qsb^{t-1})$, and abbreviate it as $\Hyb^t$ for this particular instantiation of parameters. Since $\Hyb^t$ is greedy only for one item type, its drift is easier to study, and can be shown to be proportional to $-\sqrt{\Phi(\Qsb^{t-1})}$ for $\Phi(\Qsb^{t-1})$ large enough (Lemma~\ref{lem:negative_drift_main}).

However, to be able to claim that under policy $\Rcal(\mathring{\lambdab}^t)$ the expected change in the Sum-of-Squares potential is $\Ocal(1)$, we 
require the stricter condition: $\lambdab^t  \in \Bcal_{\epsilon_0-\delta,\TV}(\bar{\lambdab})$, so that
\[ \norm{\mathring{\lambdab}^t - \bar{\lambdab}}_\TV \leq \norm{\mathring{\lambdab}^t - {\lambdab}^t}_\TV  + \norm{{\lambdab}^t - \bar{\lambdab}}_\TV  \leq  \epsilon_0.\]


{\bf The obstacle} in using the above proof approach is that $\lambdab^t$ need not be in $\Bcal_{\epsilon_0-\delta,\TV}(\bar{\lambdab})$, but the smoothed distributions $\hat{\lambdab}^t$ are. Therefore, while the following is true:
\[ \expctsub{\lambdab^t, \SoS}{\Delta^t f} \leq \expctsub{\lambdab^t, \pi^t}{\Delta^t f},\]
the right hand side of the above need not give a negative drift for a $t$ where $\Phi(\Qsb^{t-1})$ is large. Instead, we redefine $\Hyb^t:=\Hyb(\hat{\lambdab}^t,\Qsb^{t-1})$, and rewrite the right hand side as
\begin{align}
    \label{eqn:drift_decomposition}
    \expctsub{\lambdab^t, \SoS}{\Delta^t f} \leq \expctsub{\hat{\lambdab}^t, \Hyb^t}{\Delta^t f} + \left(  \expctsub{\lambdab^t, \pi^t}{\Delta^t f} -  \expctsub{\hat{\lambdab}^t, \Hyb^t}{\Delta^t f} \right).
    \end{align}
The first term above has the desired properties of being bounded, and being a large negative value when $\Phi(\Qsb^{t-1})$ is large. For the second term, we will use a trick from amortized analysis of online algorithms. We add another term $\Xi^t$ to the potential function so that $f^t = \Psi^t = \sqrt{\Phi(\Qsb^t) + \Xi^t}$. The term  $\Xi^t$ acts like a non-negative ``bank balance,'' so that whenever the second term of \eqref{eqn:drift_decomposition} is positive leading to an increase in $\Phi^t$, then $\Xi^t$ decreases by enough (i.e., we can draw on our bank balance to pay for this extra cost) so that,
\begin{align*}
    \Delta^t \Xi & \approx    \expctsub{\hat{\lambdab}^t, \Hyb^t}{\Delta^t \Phi} - \expctsub{\lambdab^t, \pi^t}{\Delta^t \Phi} \\
    \intertext{and consequently,}
    \expctsub{\lambdab^t, \SoS}{\Delta^t \Phi + \Delta^t \Xi}   &  \lessapprox  \expctsub{\hat{\lambdab}^t, \Hyb^t}{\Delta^t \Phi} .
    \end{align*}
    
A crucial component of this is a careful construction of the policy $\pi^t$ (which is only for the purposes of analysis), where we use the fact that by the smoothed $\GPG_{\delta, \tau}$ assumption, $\lambdab^t$ satisfy 
\[ \sum_{s=1}^{t-2\tau} \hat{\lambdab}^s \leq \sum_{s=\tau + 1}^t  {\lambdab}^s \leq \sum_{s=1}^t \hat{\lambdab}^s. \]
Therefore, we can \textit{couple} the probability mass of arrival at time $t$ under $\lambdab^t$ to some probability mass of arrival during interval $[t-2\tau,t]$ under $\hat{\lambdab}^s$. To be slightly more precise, the randomized policy $\pi^t$ follows some randomized action taken by policy $\Hyb^s$ during the interval $s \in [t-2\tau,t]$. This ensures that if $\expctsub{\lambdab^t, \pi^t}{\Delta^t \Phi}$ is large, then it was already paid for by $\expctsub{\hat{\lambdab}^s,\Hyb^s}{\Delta^s  \Phi}$ at an earlier time, which is part of our bank balance $\Xi^t$. One final wrinkle is that, this randomized action is taken by $\Hyb^s$ and $\pi^t$ at different times, and at different states of the virtual queues. However, since $|s-t|\leq 2\tau$, the difference in the change in $\Phi$ due to this can be bounded by $16\tau$ (Lemma~\ref{lem:time_shift}). We now proceed with the formal proof.

\subsubsection{Preliminary lemmas}
\ \\
We begin with some preliminary lemmas which are proved at the end of this Section.

The first lemma gives the expression in the change of the sum-of-squares potential for a configuration when an item is added. \begin{lemma}[Elementary Cost Functions]
\label{lem:SS_change_1}
Consider a configuration $m$  with virtual state $\Qsb_m$. Without loss of generality, let $\Ical(m) = [\ell]$, and $\sigma_m = (1,2,\ldots, \ell)$. Define $\delta_i = \Qs_{im}-\Qs_{(i-1)m}$ for $i \in \{2,\ldots, \ell\}$, and $\delta_1=\Qs_{1m}-\Qs_{\ell m}$, with the convention $\delta_{\ell+1}:=\delta_1$.
Let one unit of resource type $i$ be committed to $m$ resulting in new virtual state $\Qsb'_m$ (potentially after a virtual match). Then,
\begin{align*}
    \cost(\Qsb,i,m) := \Phi_m(\Qsb'_m) - \Phi_m(\Qsb_m) &= 2(\delta_i - \delta_{i+1}) + 2.
\end{align*}
The elementary cost is bounded as $|\cost(\Qsb,i,m)| \leq \sqrt{8 \Phi(\Qsb)}+2$.
\end{lemma}

The next lemma gives an almost sure bound on the change of the potential function $\Psi^t$. This turns out to be a bit weak for our purposes, we prove a strengthened version during the proof of Lemma~\ref{lem:drift_lemma}.
\begin{lemma}[Bounded Increment Lemma]
\label{lem:bounded increment}
Let $\Qsb$ be an arbitrary state of virtual queues, into which an item is packed to yield new virtual state $\Qsb'$. Then for $\Xi,\Xi' \geq 0$,
\[ \left| \sqrt{\Phi(\Qsb') + \Xi' } - \sqrt{\Phi(\Qsb) + \Xi } \right| \leq  \sqrt{2} + |\Xi' - \Xi|. \]
\end{lemma}

{\blue The next three lemmas prove that the expected drift of the Sum-of-Square function $\Phi$ is negative if the potential function is large. Since these lemmas only deal with $\Phi$, and not the combined Lyapunov function $\Psi^t$, the proofs of these Lemmas largely follow the tools developed by \cite{csirik2006sum}. It is also these three lemmas which suffice for the analysis in Section~\ref{sec:iid} for the simpler setting of \textit{i.i.d.} arrivals and online-queueable resources.}
\begin{lemma}
\label{lem:expected_inc}
Consider a state $\Qsb$ encountered during the execution of Algorithm~\ref{alg:SoS}, into which an item sampled from a distribution $\lambdab \in \Bcal_{\epsilon_0,\TV}(\bar{\lambdab})$ is packed according to $\Rcal(\lambdab)$ to obtain a state $\Qsb'$. Then the expected change in Sum-of-squares potential is:
\[ \expctsub{\lambdab,\Rcal(\lambdab)}{  \Phi(\Qsb')-\Phi(\Qsb) } = 2 .\]
\end{lemma}

\begin{lemma}
\label{lem:negative_drift}
Consider a state $\Qsb$ encountered during the execution of $\SoS$ such that $\Phi(\Qsb) \geq B^2$ where $B := 2n^2$. Then there exists a resource type $i^*$, such that if an item of type $i^*$ is packed by $\SoS$, the resulting virtual queue state satisfies 
\[ \Phi(\Qsb')-\Phi(\Qsb)  \leq - 2 \frac{\sqrt{\Phi(\Qsb)}}{B} .\]
\end{lemma}

As a consequence of Lemma~\ref{lem:expected_inc} and Lemma~\ref{lem:negative_drift}, the next lemma shows the existence of a Hybrid randomized+greedy policy with large negative drift, and which by definition also bounds the drift of $\SoS$.

\begin{lemma}[Drift Lemma]
\label{lem:negative_drift_main}
Consider a state $\Qsb$ such that $\Phi(\Qsb) \geq \frac{4B^2}{\delta^2}$ where $B := 2n^2$, into which an item of type sampled from $\lambdab$, $\lambdab \in \Bcal_{\epsilon_0-\delta, \TV}(\bar{\lambdab})$ has to be packed. Then, there exists a policy $\Hyb(\lambdab,\Qsb)$, such that the resulting virtual queue state satisfies 
\[ \expctsub{\lambdab,\SoS}{\Phi(\Qsb')-\Phi(\Qsb)}  \leq \expctsub{\lambdab,\Hyb(\lambdab,\Qsb)}{\Phi(\Qsb')-\Phi(\Qsb)} \leq - \frac{\delta \sqrt{ \Phi(\Qsb)}}{B}.\]
\end{lemma}

\begin{lemma}[Lipschitz Cost Lemma]
\label{lem:time_shift}
Let $\Qsb^{s}$ and $\Qsb^{t}$ be two states of the virtual queues at times $s,t$ respectively under some matching algorithm. Fix any configuration $m$, and resource type $i\in \Ical(m)$. Let $\hat{\Qsb}^s$ and $\hat{\Qsb}^t$ be the states of the virtual queues after committing $i$ to configuration $m$ in states $\Qsb^s$ and $\Qsb^t$, respectively. Then
\[  \left| \left(\Phi_m(\hat{\Qsb}^t_m) - \Phi_m({\Qsb}^t_m) \right) -\left(\Phi_m(\hat{\Qsb}^s_m) - \Phi_m({\Qsb}^s_m) \right) \right| \leq 8|s-t|.  \]
\end{lemma}

\subsubsection{Formal Proof}
\ \\
As we mentioned in the proof sketch, we bound the (amortized) drift in $\Phi + \Xi$ at time $t$ by the drift under the hybrid policy $\Hyb(\hat{\lambdab}^t,\Qsb^{t-1})$, and use a cleverly constructed sequence of policies $\{\pi^t\}$ \textit{coupled} to the sequence $\Hyb(\hat{\lambdab}^t,\Qsb^{t-1})$ to upper bound the drift of $\SoS$. We first explain the construction of this sequence $\{\pi^t\}$.

\vspace{3mm}

{\bf Construction of $\pi^t$:} We begin by defining a sequence of functions for each $i \in \Ical$ and $m \in \Mcal_+$ based on the actions of $\Hyb^t:=\Hyb(\hat{\lambdab}^t, \Qsb^{t-1})$. Note that $\Hyb^t$ is $\Fcal^{t-1}$-measurable. Let the probability that $\Hyb^t$ commits resource type $i$ to configuration $m$ be $\hat{p}^t_{im}$. Based on these, we define the piecewise constant functions $\hat{q}_{im}:[0,T] \to [0,1]$ as follows:
\begin{align}
    \hat{q}_{im}(s) &= \hat{\lambda}_i^{\ceil{s}} \cdot \hat{p}^{\ceil{s}}_{im} , \qquad \forall \ s\in [0,T].
\end{align}
The following helper functions will be useful for defining $\pi^t$. We use $\hat{\Lambda}_i(t):[0,T]\to [0,T]$ to denote the cumulative (fractional) expected number of arrivals of type $i$ by time $t$ under the sequence of distributions $\{\hat{\lambdab}^t\}$ (here $[0,T]$ is the continuous interval). By $\Lambda_i(t):[\tau,T]\to [0,T]$ we denote the cumulative expected number of arrivals of type $i$ in the interval $\{\tau+1, \ldots , t\}$ according to the sequence $\{\lambdab^t\}$:
\begin{align}
    \hat{\Lambda}_i(t) &= \sum_{s=1}^{\floor{t}} \hat{\lambda}^s_i + (t - \floor{t}) \hat{\lambda}_i^{\ceil{t}}, \qquad \forall t \in [0,T], \\
    {\Lambda}_i(t) &= \sum_{s=\tau+1}^{t} {\lambda}^s_i, \qquad \forall t \in \{\tau+1,\ldots,T\}. 
\end{align}
By the smoothed $\GPG_{\delta,\tau}$ condition we have,
\begin{align}
\label{eqn:time_shift_bound}
     t-2\tau \leq  \hat{\Lambda}_i^{-1}(t) \leq t. 
\end{align}

The policy $\pi^t$ is only defined for $t \in \{\tau+1,\ldots, T\}$ (the potential is bounded until time $\tau$, and our drift analysis only begins at time $\tau+1$). Recall that the policy $\pi^t$ sees a resource with type sampled from $\lambdab^t$. The probability that policy $\pi^t$ commits an item of type $i$ to configuration $m$ is given by 
\begin{align}
    p^t_{im} &= \frac{1}{\lambda^t_i} \int_{s = \hat{\Lambda}^{-1}_i(\Lambda_i(t-1))}^{{s = \hat{\Lambda}^{-1}_i(\Lambda_i(t))}} \hat{q}_{im}(s) ds.
\end{align}
Intuitively, policy $\pi^t$ looks at the cumulative probability with which $\Hyb$ sends resource $i$ to configuration $m$ during the (real-valued) interval  $\left[ \hat{\Lambda}^{-1}_i(\Lambda_i(t)),  \hat{\Lambda}^{-1}_i(\Lambda_i(t-1)) \right]$. In this sense, the randomized action of the $\pi^t$ for resource type $i$ is coupled with the randomized action of $\Hyb$ for resource type $i$ during this interval. The intervals for different resource types need not be (and in general will not be) the same.

\vspace{3mm}

{\bf Relating cost of $\pi^t$ and $\Hyb$:} By definition, the expected change of the potential under policy $\pi^t$ at time $t$ is
\[ \expctsub{\lambdab^t, \pi^t}{\Delta^t \Phi} = \sum_{i,m} p^t_{im} \cdot \cost(\Qsb^{t-1},i,m).\]
We define the cost incurred by $\Hyb^s$ for the actions contributing to $\pi^t$ as:
\begin{align}
\label{eqn:Hybrid_pi_cost}
    \cost_{\Hyb,t} &= \sum_{i} \int_{s = \hat{\Lambda}^{-1}_i(\Lambda_i(t-1))}^{{s = \hat{\Lambda}^{-1}_i(\Lambda_i(t))}} \sum_{m \in \Mcal_+} \cost(\Qsb^{s-1},i,m) \hat{q}_{im}(s) ds.
\end{align} 
In words, for every $i$, if the $\pi^t$ policy incorporates the action taken by Hybrid at time $s \leq t$, then $\cost_{\Hyb,t}$ incorporates the cost paid by $\Hyb^s$ due to item $i$ (in proportion to amount of contribution of time $s$ to $q^{t}_{im}$). The key idea is that due to this coupling of $\pi^t$ and $\Hyb$, \eqref{eqn:time_shift_bound} and Lemma~\ref{lem:time_shift} yield the key inequality:
\begin{align}
\label{eqn:bound_cost_pit}
    \expctsub{\lambdab^t, \pi^t}{\Delta^t \Phi} \leq \cost_{\Hyb,t} + 16 \tau.
\end{align}

\vspace{3mm}

{\bf Amortized Lyapunov Drift analysis:} We first define the compensator $\Xi^t$, and the Lyapunov function $\Psi^t$ for Lemma~\ref{lem:drift_lemma}:
\begin{align}
\label{eqn:def_Xi}
    \Xi^t & := \sum_{s=1}^t \expctsub{\hat{\lambdab}^s,\Hyb^s}{\Delta^s \Phi} - \sum_{s=\tau+1}^t \cost_{\Hyb,s} + \tau \max_{ t-2\tau \leq s \leq t } \left( \sqrt{ 8 \Phi(\Qsb^s) }+ 2 \right),\\
    \Psi^t &= \sqrt{\Phi(\Qsb^t) + \Xi^t}.
    \label{eqn:def_Psi}
\end{align}
Our goal is to show that with the above definition of $\Xi^t$, \textit{(i)} $\Xi^t$ is non-negative for all $t$ (so that $\Psi^t$ bounds $\Phi^t$), \textit{(ii)} $\Xi^t$ is not too large (so that the bound is not loose), and \textit{(iii)} that the bounded increment and expected decrease conditions from Lemma~\ref{lem:drift_lemma} hold for $\Psi^t$.

To show non-negativity, focus on the first part of $\Xi^t$:
\[ \Xi_1^t := \sum_{s=1}^t \expctsub{\hat{\lambdab}^s,\Hyb^s}{\Delta^s \Phi} - \sum_{s=\tau+1}^t \cost_{\Hyb,s}. \]
Both the first and second term in the summation involve the elementary cost functions for the hybrid policy; the first sum is the cost in the interval $\{1,\ldots,t\}$, and the second term involves the coupling used for the policies $\pi^{\tau+1},\ldots, \pi^t$, and therefore omits for each $i$, some time interval in $[0,t]$. Crucially, by \eqref{eqn:time_shift_bound}, the support of the missing mass is entirely on $\{t-2\tau,\ldots,t\}$, and the total missing mass is bounded by $\tau$. By Lemma~\ref{lem:SS_change_1}, the absolute value of the elementary cost for any action at time $s$ is bounded by $\sqrt{8\Phi(\Qsb^s)}+2$, and hence $\max_{t-2\tau \leq s \leq t}\sqrt{8\Phi(\Qsb^s)}+2$ uniformly upper bounds the norm of missing elementary cost function mass in the summation.

To show that the $\Xi^t$ is small, the above argument also immediately implies:
\begin{align}
\label{eqn:bound_Xi}
    \Xi^t & \leq 2 \tau \max_{ t-2\tau \leq s \leq t } \left( \sqrt{ 8 \Phi(\Qsb^s) }+ 2 \right),
    \intertext{and applying Lemma~\ref{lem:sqrt_increment} (proved at the end of this subsection), we can further bound the above in terms of $\Phi(\Qsb^t)$ as}
    \Xi^t & \leq 2 \tau \left( \sqrt{ 8 \Phi(\Qsb^t) }+ 8\tau + 2 \ln (1+4\tau) \right).
    \label{eqn:bound_Xi_2}
\end{align}

We now prove a drift bound for $\Phi(\Qsb^t) + \Xi^t$. Since $\SoS$ minimizes the drift in $\Phi$, the change in $\Phi$ under $\SoS$ can be upper bounded by the drift under $\pi^t$:
\begin{align}
\nonumber
    \expctsub{\lambdab^t, \SoS}{\Delta^t \Phi + \Delta^t \Xi} & \leq \expctsub{\lambdab^t, \pi^t}{\Delta^t \Phi} + \expctsub{\lambdab^t, \SoS}{\Delta^t \Xi} \\
\nonumber
    & \stackrel{\eqref{eqn:bound_cost_pit}}{\leq} \cost_{\Hyb,t} + 16 \tau + \expctsub{\lambdab^t, \SoS}{\Delta^t \Xi},\\
    \intertext{adding and subtracting $\expctsub{\hat{\lambdab}^t,\Hyb^t}{\Delta^t \Phi}$:}
\nonumber
    & = \expctsub{\hat{\lambdab}^t,\Hyb^t}{\Delta^t \Phi} + 16 \tau +    \cost_{\Hyb,t} - \expctsub{\hat{\lambdab}^t,\Hyb^t}{\Delta^t \Phi} + \expctsub{\lambdab^t, \SoS}{\Delta^t \Xi},\\
    \intertext{using \eqref{eqn:def_Xi} to expand $\expctsub{\lambdab^t, \SoS}{\Delta^t \Xi}$:}
\nonumber
    &= \expctsub{\hat{\lambdab}^t,\Hyb^t}{\Delta^t \Phi} + 16 \tau +    \cost_{\Hyb,t} - \expctsub{\hat{\lambdab}^t,\Hyb^t}{\Delta^t \Phi} \\
\nonumber
    & \qquad + \expctsub{\hat{\lambdab}^t,\Hyb^t}{\Delta^t \Phi} -  \cost_{\Hyb,t} + \tau \left( \max_{ t-2\tau \leq s \leq t }  \sqrt{ 8 \Phi(\Qsb^s) } - \max_{ t-2\tau-1 \leq s \leq t-1 }  \sqrt{ 8 \Phi(\Qsb^s) } \right)\\
\nonumber
    &= \expctsub{\hat{\lambdab}^t,\Hyb^t}{\Delta^t \Phi} + 16 \tau +     \tau \left( \max_{ t-2\tau \leq s \leq t }  \sqrt{ 8 \Phi(\Qsb^s) } - \max_{ t-2\tau-1 \leq s \leq t-1 }  \sqrt{ 8 \Phi(\Qsb^s) } \right), \\
    \intertext{applying Lemma~\ref{lem:sqrt_increment} to bound the term in parenthesis above,}
\nonumber
    &\leq \expctsub{\hat{\lambdab}^t,\Hyb^t}{\Delta^t \Phi} + 16 \tau + \tau \left( 8\tau+4 + 2 \log ( 3 + 4 \tau)  \right) \\
    &\leq \expctsub{\hat{\lambdab}^t,\Hyb^t}{\Delta^t \Phi} +   8\tau^2 + 2 \tau\log ( 3 + 4 \tau) + 20 \tau .
\label{eqn:initial_bound_Psi}
\end{align}

\vspace{3mm}

{\bf Bounded increment for $\Psi^t$:} Now we turn to proving the bounded increment property for $\Psi^t$, which improves upon Lemma~\ref{lem:bounded increment}. Lemma~\ref{lem:bounded increment} has $\Xi'-\Xi$ on the right hand side, which is in fact not bounded since it involves $\expctsub{\hat{\lambdab}^t,\Hyb^t}{\Delta^t \Phi} -  \cost_{\Hyb,t}$ which is bounded by $2\tau \sqrt{8 \Phi(\Qsb)^t}$.

First note that, we can upper bound
\begin{align*}
    \Xi^t - \Xi^{t-1} &= \expctsub{\hat{\lambdab}^t,\Hyb^t}{\Delta^t \Phi} -  \cost_{\Hyb,t} +  \tau \left( \max_{ t-2\tau \leq s \leq t }  \sqrt{ 8 \Phi(\Qsb^s) } - \max_{ t-2\tau-1 \leq s \leq t-1 }  \sqrt{ 8 \Phi(\Qsb^s) } \right) \\
    & \leq 2\tau \sqrt{8 \Phi(\Qsb)^t} + 8 \tau^2 + 2\tau \log(3+4\tau) + 4 \tau \\
    & \leq 2\tau \sqrt{8 \Phi(\Qsb)^{t-1}} + 8 \tau^2 + 2\tau \log(3+4\tau) + 4 \tau + 8.
\end{align*}
Now we proceed through the steps of proof of Lemma~\ref{lem:bounded increment}:
First we assume $\Phi(\Qsb^t) \geq \Phi(\Qsb^{t-1})$. We have, for some $x,y \geq 0$,
\[ \Phi(\Qsb^t) \leq \Phi(\Qsb^{t-1}) + (x+1)^2 +(y+1)^2 - x^2-y^2.\] 
Using this,
\begin{align*}
 \sqrt{\Phi(\Qsb^t) + \Xi^t } - \sqrt{\Phi(\Qsb^{t-1}) + \Xi^{t-1} }  &\leq  \sqrt{ \Phi(\Qsb^t) + (x+1)^2+(y+1)^2 - x^2 - y^2 + \Xi^t } - \sqrt{\Phi(\Qsb^{t-1}) + \Xi^{t-1} } \\
 &= \frac{2x+2y+2 + \Xi^t-\Xi^{t-1}}{\sqrt{\Phi(\Qsb) +(x+1)^2+(y+1)^2 - x^2 - y^2+ \Xi^t } + \sqrt{\Phi(\Qsb^{t-1}) + \Xi^{t-1} } } \\
 & \leq \frac{2x+2y+2}{\sqrt{(x+1)^2+(y+1)^2} + \sqrt{x^2 + y^2} } 
 +  \frac{\Xi^t-\Xi^{t-1}}{\sqrt{2} + \sqrt{\Phi(\Qsb^{t-1})} }\\
 & \leq \sqrt{2} + \frac{2\tau \sqrt{8 \Phi(\Qsb)^{t-1}} + 8 \tau^2 + 2\tau \log(3+4\tau) + 4 \tau + 8}{\sqrt{2} + \sqrt{\Phi(\Qsb^{t-1})}} \\
 & \leq \sqrt{2} + 2\tau\sqrt{8} + \frac{8}{\sqrt{2}} \tau^2 + 2\tau \log(3+4\tau) + \frac{4}{\sqrt{2}} \tau + \frac{8}{\sqrt{2}}\\
 & \leq 6 \tau^2 + 9 \tau + 2\tau \log(3+4\tau) + 8\\
 & =: K_1(\tau).
\end{align*}

\vspace{3mm}

{\bf Expected Decrease for $\Psi^t$: } We use the following three facts established previously,
\begin{enumerate}[label=(\roman*)]
\item From \eqref{eqn:initial_bound_Psi},
\begin{align*}
    \expctsub{\lambdab^t, \SoS}{\Delta^t \Phi + \Delta^t \Xi } & \leq \expctsub{\hat{\lambdab}^t,\Hyb^t}{\Delta^t \Phi} +  8\tau^2 + 2 \tau\log ( 3 + 4 \tau) + 20 \tau.
\end{align*}
\item From Lemma~\ref{lem:negative_drift_main}, if $\Phi(\Qsb^{t-1}) \geq \frac{4B^2}{\delta^2}$, where $B=2n^2$, then, 
\[ \expctsub{\hat{\lambdab}^t,\Hyb^t}{\Delta^t \Phi} \leq - \frac{\delta \sqrt{ \Phi(\Qsb^{t-1})}}{B}. \]
From (i) and (ii), if $\Phi(\Qsb^{t-1}) \geq c_1(\tau, \delta)$
where
\begin{align}
c_1(\tau, \delta) := \frac{4B^2 ( 8\tau^2 + 2 \tau\log ( 3 + 4 \tau) + 20 \tau + 1)^2 }{\delta^2},
\label{eqn:c1}
\end{align} 

then 
\[    \expctsub{\lambdab^t, \SoS}{\Delta^t \Phi + \Delta^t \Xi } \leq -\frac{\delta\sqrt{\Phi(\Qsb^{t-1})}}{2B}. \]
\item From \eqref{eqn:bound_Xi_2},
\[    \Xi^{t-1} \leq 2 \tau \left( \sqrt{ 8 \Phi(\Qsb^{t-1}) }+ 8\tau + 2 \ln (1+4\tau) \right),\]
and therefore if $\Phi(\Qsb^{t-1}) \geq c_2(\tau)$ where 
\begin{align}
    c_2(\tau) := 32 \tau^2 + 4\tau(8\tau + 2 \ln (1+4\tau) ),
    \label{eqn:c2}
\end{align} 
is large enough so that 
\[ c_2(\tau) \geq  2 \tau \left( \sqrt{ 8 c_2(\tau)}+ 8\tau + 2 \ln (1+4\tau) \right) ,\]
then $(\Psi^{t-1})^2 = \Phi(\Qsb^{t-1})+\Xi^{t-1} \leq 2 \Phi(\Qsb^{t-1})$.
\end{enumerate}
Using the inequality (Lemma 3.6 of \cite{csirik2006sum}), for $x,y \geq 0$,
\[ x - y \leq \frac{x^2-y^2}{2y},\]
substituting $x = \Psi^t, y = \Psi^{t-1}$, when $\Phi(\Qsb^{t-1}) \geq \max\{ c_1(\tau, \delta), c_2(\tau)\}$,
\begin{align*}
    \expct{\Delta^t \Psi} \leq \frac{\expct{\Delta^t \Phi + \Delta^t \Xi}}{2\Psi^{t-1}} \leq \frac{-\delta \sqrt{\Phi(\Qsb^{t-1})}/2B}{2\sqrt{2\Phi(\Qsb^{t-1})}} \leq -\frac{\delta }{4\sqrt{2}B}.
\end{align*}
\hfill \BlackBox

\begin{lemma}
\label{lem:sqrt_increment}
Let $a_0 \geq 0$, and consider the series $a_t = a_{t-1} + \sqrt{8a_{t-1}}+2$. Then,
\[  \sqrt{8a_t} - \sqrt{8a_0} \leq 4t + 2 \log(1+2t).  \]
\end{lemma}
\proof{Proof:}
We can upper bound $a_t$ by $x(t)$ which solves the following differential equation:
\[ \frac{dx}{dt} = \sqrt{8x}+2 \]
with initial condition $x(0) = a_0$. The solution of this differential equation is given by
\[  \sqrt{8x(t)} - \sqrt{8x(0)} = 4t + 2 \log \frac{\sqrt{8x(t)}+2}{\sqrt{8x(0)}+2}. \]
To upper bound the ratio inside the $\log$, we consider the solution of the differential equation with initial condition $x(0)=0$, which gives an upper bound on $\sqrt{8x(t)}$ of $4t$. Therefore, we get the bound $\frac{\sqrt{8x(t)}+2}{\sqrt{8x(0)}+2} \leq \frac{4\cdot t + 2}{4\cdot 0 + 2} = 1+2t$ for all $x(0)\geq 0$. Substituting this upper bound into the solution of the differential equation
\[ \sqrt{8x(t)} - \sqrt{8x(0)} \leq 4t + 2\log(1+2t). \]
\hfill \BlackBox
\endproof

\subsubsection{Proofs of supporting lemmas}
\ \\
\begin{proofof}{Lemma~\ref{lem:SS_change_1}}
The proof is a straightforward computation from the definition \eqref{eqn:SS}. If an resource of type $i$ arrives and no virtual match is complete then $\Qs'_{im} = \Qs_{im}+1$, and the rest of the virtual queue lengths are unchanged. Therefore, $\delta_i$ increases by 1, and $\delta_{i+1}$ decreases by 1. Since 
\[ \Phi_m(\Qsb_m) = \sum_{k=1}^\ell \delta_k^2, \]
the overall change in $\Phi_m$ is $2(\delta_i-\delta_{i+1})+2$.

Now consider the case that type $i$ resource arrives and a virtual match happens. In this case $\Qs'_{im}=\Qs_{im}$, and the rest of the virtual queue lengths decrease by $1$. Net effect is still that $\delta_i$ increases by 1, $\delta_{i+1}$ decreases by 1, and the rest of the $\delta_k$ are unchanged.

Since $|\delta_i|, |\delta_{i+1}| \leq \Phi(\Qsb) \leq \sqrt{|\delta_i|^2+\delta_{i+1}^2}$, it follows that $|\delta_i|+|\delta_{i+1}| \leq \sqrt{2\Phi(\Qsb)}$, from which the bound on $\cost(\Qsb,i,m)$ follows.
\end{proofof}

\vspace{3mm}

\begin{proofof}{Lemma~\ref{lem:bounded increment}}
Let $m$ be the configuration into which type $i$ resource is added. Consider the case in which $\Phi_m(\Qsb'_m) \geq \Phi_m(\Qsb_m)$. As described in the proof of Lemma~\ref{lem:SS_change_1}, in this case $\delta_i$ increases and $\delta_{i+1}$ decreases. In the worst case, $\delta_i$ was positive, and $\delta_{i+1}$ was negative, so both changes lead to an increase in $\Phi_m$. Let the original value of $\delta_i=x$ and $\delta_{i+1}=-y$ for $x,y \geq 0$. Then,
\[ \Phi_m(\Qsb'_m) \leq \Phi_m(\Qsb_m) + (x+1)^2 +(y+1)^2 - x^2-y^2.\]
Now, following the proof of Lemma 3.5 of \cite{csirik2006sum}:
\begin{align*}
 \sqrt{\Phi(\Qsb') + \Xi' } - \sqrt{\Phi(\Qsb) + \Xi }  &\leq  \sqrt{ \Phi(\Qsb) + (x+1)^2+(y+1)^2 - x^2 - y^2 + \Xi' } - \sqrt{\Phi(\Qsb) + \Xi } \\
 &= \frac{2x+2y+2 + \Xi'-\Xi}{\sqrt{\Phi(\Qsb) +(x+1)^2+(y+1)^2 - x^2 - y^2+ \Xi' } + \sqrt{\Phi(\Qsb) + \Xi } } \\
 & \leq \frac{2x+2y+2 + \Xi'-\Xi}{\sqrt{(x+1)^2+(y+1)^2} + \sqrt{x^2 + y^2} } \\
 & = \frac{2x+2y+2 + \Xi'-\Xi}{\sqrt{x^2+y^2+2x+2y+2} + \sqrt{x^2+y^2} } \\
 \intertext{and letting $z=x+y$ and minimizing $x^2+y^2$ to get $x=y=z/2$, we further upper bound the above by,}
 & \leq \frac{2z+2 + \Xi'-\Xi}{\sqrt{\frac{z^2}{2}+2z+2}+\sqrt{\frac{z^2}{2}}} \\
& = \sqrt{2}\frac{2z+2 + \Xi'-\Xi}{(z+2)+z} \\
 & \leq \sqrt{2} + |\Xi'-\Xi|.
\end{align*}
The case $\Phi_m(\Qsb'_m)\leq \Phi_m(\Qsb_m)$ is handled similarly.
\end{proofof}

\vspace{3mm}

\begin{proofof}{Lemma~\ref{lem:expected_inc}}
It suffices to prove the result for the case where the configuration $m$ to which an item has to be committed is fixed, and the probability that the type of arriving item $I=i$ is $p_i = M_{im}/\sum_{i'}M_{i'm}=1/\ell_m$. We can then view the distribution $\lambdab$ as first sampling a configuration $m$ according to the solution of $\SPP(\lambdab)$, and then sampling the type of the item $i$ as in the previous sentence. Since by Lemma~\ref{lem:SS_change_1}, the change in $\Phi_m$ is $2(\delta_i-\delta_{i+1})+2$ when a resource a type $i$ arrives, the expected change if a resource $i \in \Ical(m)$ arrives uniformly at random,
\[ \expct{\Phi_m(\Qsb'_m) - \Phi_m(\Qsb_m)} = \frac{1}{\ell_m} \sum_{i=1}^{\ell_m} \left( 2(\delta_i - \delta_{i+1}) +2 \right) = 2.\]
\end{proofof}

\vspace{3mm}

\begin{proofof}{Lemma~\ref{lem:negative_drift}}
Let $\Phi := \sum_m \Phi_m(\Qsb_m)$. Since  $|\Mcal_+| = n$, there must exist a configuration $m$ such that $\Phi_m(\Qsb_m) \geq \Phi/n$. Using the notation and setup from Lemma~\ref{lem:SS_change_1}, since $\Phi_m(\Qsb_m) = \sum_{k=1}^\ell \delta_k^2$, there must exist a $j$ such that $|\delta_j| \geq \sqrt{\Phi}/n$, and let $i = \argmax_{k \leq j-1} \delta_{k+1}-\delta_k$. Since $\sum_{k} \delta_k = 0$, it must be the case that $\delta_{i+1}-\delta_{i} \geq \frac{\sqrt{\Phi}}{n^2}$. By Lemma~\ref{lem:SS_change_1}, if the arriving item is of type $i^*=i$, then the change in $\Phi_m$ is
\[ \Phi_m(\Qsb'_m)-\Phi_m(\Qsb_m) = -2(\delta_{i+1}-\delta_i)+2 \leq - \frac{2\sqrt{\Phi}}{n^2} + 2 \leq -4 \frac{\sqrt{\Phi}}{B}+2 \leq -2 \frac{\sqrt{\Phi}}{B} \]
where the second inequality follows from $B=2n^2$, and the next from the assumption $\Phi \geq B^2$.
\end{proofof}

\vspace{3mm}

\begin{proofof}{Lemma~\ref{lem:negative_drift_main}}
By Lemma~\ref{lem:negative_drift}, there is a resource type $i^*$ which if packed into $\Qsb$ according to the $\SoS$ algorithm leads to a change in sum-of-squares potential of at least $-2 \frac{\sqrt{\Phi}}{B}$. 
We decompose $\lambdab$ into 
\[ \lambdab = \delta \eb_{i^*} + (1-\delta) \underbrace{\left( \frac{\lambdab - \eb_{i^*}}{1-\delta} \right)}_{=: \mathring{\lambdab}}.\]
The policy $\Hyb(\lambdab, \Qsb)$ is defined to be the mixture of $\Rcal(\mathring{\lambdab})$ and $\SoS$: If the item type is $i \neq i^*$, $\Hyb$ picks the distribution to commit $i$ to based on the randomized policy $\Rcal(\mathring{\lambdab})$. If the item type is $i=i^*$, then $\Hyb$ follows $\SoS$ with probability $\delta/\lambda_i$, and $\Rcal(\mathring{\lambdab})$ otherwise.

Then since $\SoS$ greedily minimizes $\Phi$,
\[ \expctsub{\lambdab, \SoS}{\Phi(\Qsb')-\Phi(\Qsb)} \leq (1-\delta) \expctsub{\mathring{\lambdab}, \Rcal(\mathring{\lambdab})}{\Phi(\Qsb')-\Phi(\Qsb)} + \delta \expctsub{\eb_{i^*}, \SoS}{\Phi(\Qsb')-\Phi(\Qsb)}.
\]
We have $\norm{\mathring{\lambdab} - \bar{\lambdab}}_\TV \leq \norm{\mathring{\lambdab} - {\lambdab}}_\TV  + \norm{{\lambdab} - \bar{\lambdab}}_\TV  \leq  \epsilon_0$, and therefore $\expctsub{\mathring{\lambdab}, \Rcal(\mathring{\lambdab})}{\Phi(\Qsb')-\Phi(\Qsb)} \leq 2$ by Lemma~\ref{lem:expected_inc}. Therefore,
\begin{align*}
\expctsub{\lambdab, \SoS}{\Phi(\Qsb')-\Phi(\Qsb)} & \leq (1-\delta) 2 - \delta \frac{2\sqrt{\Phi(\Qsb)}}{B} \\
&\leq 2 - 2\delta \frac{\sqrt{\Phi(\Qsb)}}{B} \\
& \leq - \delta \frac{\sqrt{\Phi(\Qsb)}}{B}
\end{align*}
when $\Phi(\Qsb) \geq \frac{4B^2}{\delta^2}$.
\end{proofof}

\vspace{3mm}

\begin{proofof}{Lemma~\ref{lem:time_shift}}
Recall the setup and notation of Lemma~\ref{lem:SS_change_1}. Let $s < t$, and $\tau = t-s$. Since at most $\tau$ items have been packed in  configuration $m$ during the times steps $[s+1,t]$, each $\Qs_{km}$ could have changed by at most $\tau$, and therefore each $\delta_i$ can change by at most $2\tau$, and $2(\delta_i -\delta_{i+1})$ can change by at most $8\tau$.
\end{proofof}

\subsection{Proof of Lemma~\ref{lem:LB_regret}}

First consider the case where for some $t \in [T/2,T]$, $\expct{N^t_1} \geq 2c \log T$. We can lower bound the regret as,
\[ \OPT(\Ab^t) - \left( 5 X^t_1 + X^t_2 \right) \geq \ind\{ A^t_2 + A^t_3 \geq A^t_1\} N^t_1, \]
or,
\[ \expct{OPT(\Ab^t) - \left( 5 X^t_1 + X^t_2 \right)} \geq \prob{ A^t_2 + A^t_3 \geq A^t_1} \expct{N^t_1 | A^t_2 + A^t_3 \geq A^t_1}. \]
Since $A^t_2 + A^t_3 - A^t_1$ can be written as the sum of $t$ i.i.d. random variables with take the value $+1$ with probability $0.64$ and $-1$ with probability $0.36$, a simple application of Hoeffding bound gives
\[ \prob{A^t_1 - A^t_2 - A^t_3 \geq 0} \leq e^{ -\frac{2(0.28t)^2}{4t} } \leq e^{ - T/60}. \]
Also, 
\begin{align*}
     \prob{ A^t_2 + A^t_3 \geq A^t_1}  \expct{N^t_1 | A^t_2 + A^t_3 \geq A^t_1} &= \expct{N^t_1} -  \expct{N^t_1 | A^t_2 + A^t_3 < A^t_1}\cdot \prob{A^t_2 + A^t_3 < A^t_1} \\
     & \geq 2c \log T - T \cdot \prob{A^t_1 - A^t_2 - A^t_3 \geq 0} \\
     & \geq 2c \log T - T e^{- \frac{T}{60}}.
\end{align*}
For $T \geq T_0$ large enough, we get $ 2c \log T - T e^{- \frac{T}{60}} \geq  c\log T$.

Next, consider the case that for all $t \in [T/2, T]$, $\expct{N_1^t} \leq \frac{c}{4} \log T$ (where we will choose the constant $c$ later). We first create $K = \frac{T}{2c\log T}-1$ disjoint sub-intervals of $[T/2+1,T]$ where the $k$th subinterval is $[\tau_k +1,\tau_k+ c \log T]$ where  $\tau_k = T/2 + (k-1)\cdot c \log T$. Let $\Ecal_k$ denote the event that all the arrivals in $k$th sub-interval are of resource type 2, and let $\Gcal_k$ denote the event $\{N_1^{\tau_k}\} \leq \frac{c}{2} \log T$. Because $N_1^t$ is adapted to the arrivals until time $t$, $\Ecal_k$ and $\Gcal_k$ are independent for all $k$. Further, $\Ecal_k$ are \textit{i.i.d.}. Let $\alpha = \prob{\Ecal_k}$, and let $\beta_k = \prob{\Gcal_k}$. Since by our assumption, $\expct{N_1^{t}} \leq \frac{c}{4} \log T$, we have $\prob{N_1^t \geq \frac{c}{2} \log T} \leq \frac{1}{2}$, or $\prob{N_1^t \leq \frac{c}{2} \log T} \geq 1/2$. In other words,
\[ \beta_k \geq \frac{1}{2} \mbox{ for all } k.\]
Further $\prob{\Ecal_k} = \frac{1}{T^{-c\log 0.32}}$. Let $c$ be such that $- c\log 0.32 = 0.5$, so that 
\[ \alpha = \frac{1}{\sqrt{T}}.\]
Our goal is to show that $\prob{\cup_k \Ecal_k \Gcal_k}$ is at least a constant, since on this event $\frac{c}{2}\log T$ arrivals of type 2 resources are lost due to insufficient queue of type 1 resource, leading to a regret of $\frac{5c}{2}\log T$ at time $T$. 

Let $B = \sum_{k} \Ecal_k \Gcal_k$ count the number of such events. We will use the second moment method to bound $\prob{B \geq 1}$ which says that
\[ \prob{B > 0} \geq \frac{\expct{B}^2}{\expct{B^2}}. \]
In our case, since $B$ is non-negative and integer valued, $\prob{B > 0} = \prob{B \geq 1}$.

We have 
\begin{align*}
    \expct{B} &= \expct{\sum_k \Ecal_k \Gcal_k} = \sum_{k} \expct{\Ecal_k \Gcal_k} = \sum_{k} \expct{\Ecal_k}\expct{\Gcal_k} = \alpha \cdot \sum_{k} \beta_k. 
\end{align*}
Next,
\begin{align*}
    \expct{B^2} &= \sum_{k} \expct{\Ecal_k \Gcal_k} + 2 \sum_{k < \ell} \expct{\Ecal_k \Ecal_\ell \Gcal_k \Gcal_\ell} \\
    & \leq \sum_{k} \expct{\Ecal_k} + 2 \sum_{k < \ell} \expct{\Ecal_k \Ecal_\ell} \\
    & = K \alpha + K(K-1) \alpha^2.
\end{align*}
Finally,
\[ \prob{B \geq 1} \geq \frac{\expct{B}^2}{\expct{B^2}} \geq \frac{K^2\alpha^2/4}{ K^2 \alpha^2 + K\alpha } \geq 1/8, \]
where in the last step we have used $K\alpha \geq 1$ for our choice of $c$, and $T \geq T_0$ large enough.

\end{APPENDICES}

%
%







\end{document}